\documentclass[11pt]{article}
\usepackage[margin=2.5cm]{geometry}
\usepackage[utf8]{inputenc}
\usepackage{amsmath,amsthm}              
\usepackage{hyperref}
\usepackage{graphicx}
\usepackage{colortbl}
\usepackage{multirow}
\usepackage{booktabs}
\usepackage{subcaption}
\usepackage{array}
\usepackage{etoolbox}
\usepackage{lineno}
\usepackage{authblk}
\usepackage{todonotes}
\usepackage{lineno}
\usepackage{float}
\usepackage{graphicx}
\usepackage{tcolorbox}
\usepackage{bm}
\usepackage{mathtools}
\usepackage{commath}
\usepackage{amssymb}
\usepackage{appendix}
\usepackage[detect-none]{siunitx}
\usepackage[sort&compress,numbers]{natbib}

\usepackage{array}
\newcolumntype{L}[1]{>{\raggedright\let\newline\\\arraybackslash\hspace{0pt}}m{#1}}
\newcolumntype{C}[1]{>{\centering\let\newline\\\arraybackslash\hspace{0pt}}m{#1}}
\newcolumntype{R}[1]{>{\raggedleft\let\newline\\\arraybackslash\hspace{0pt}}m{#1}}

\newtheorem{theorem}{Theorem}[section]

\def\r0{\mathcal{R}_0}

\newcommand{\ie}{\textit{i}.\textit{e}.}

\title{\textbf{Indication of demographic and epidemiological data on early-stage risks and long-term epidemic burden}}
\author[1,2]{Sicong Zhang\thanks{sicongzhang@bnbu.edu.cn}}
\author[1]{Mingyuan Wu\thanks{t330033036@mail.bnbu.edu.cn}}
\author[1]{Jian Yang\thanks{t330031315@mail.bnbu.edu.cn}}
\author[3]{Jianpeng Xiao\thanks{jpengx@163.com}}
\author[1,2]{Fan Bai\thanks{Corresponding author. Email: baifan.ubc@gmail.com}}

\affil[1]{Department of Mathematical Sciences, \protect \\
	Faculty of Science and Technology, \protect\\ Beijing Normal-Hong Kong Baptist University, \protect\\ Zhuhai, Guangdong, China}  

\affil[2]{Guangdong Provincial/Zhuhai Key Laboratory of Interdisciplinary Research and Application
for Data Science, \protect\\ Beijing Normal-Hong Kong Baptist University, \protect\\ Zhuhai 519087, China}

\affil[3]{Guangdong Provincial Institute of Public Health, \protect \\
Guangdong Provincial Center for Disease Control and Prevention, \protect \\
Guangzhou 511430, China}

\begin{document}

\maketitle
\begin{abstract}
This study investigates the relationship between age-specific cumulative infection burden and the probability of triggering a major outbreak in heterogeneous populations. We develop two modeling frameworks that incorporate empirical age-stratified contact matrices: a deterministic framework for deriving the final size relation and a stochastic framework based on a Continuous-time Markov chain epidemic model, for which a multi-type branching-process approximation is used to characterize early-stage outbreak risk. The framework is illustrated using empirical social contact data from Guangzhou, Foshan, and Zhuhai, China, under two disease-motivated scenarios representing COVID-19 and hand, foot, and mouth disease (HFMD). The illustrative analyses indicate that the relationship between cumulative infection burden and early-stage outbreak risk depends on both epidemiological characteristics and population contact patterns. In particular, age groups with the highest final attack rates do not necessarily have the highest probability of triggering a major outbreak. These findings highlight the complementary nature of the two measures and the value of considering both when assessing age-specific epidemic risks in practice.
\end{abstract}

\medskip

\noindent {Keywords:} Epidemiology, Population heterogeneity, Contact pattern, Probability of a major outbreak, Final epidemic attack rate.
\\

\medskip



\medspace

\section{Introduction}
Age-specific contact patterns and epidemiological characteristics play important roles in shaping the spread of infectious diseases. Individuals in different age groups may differ in their contact patterns, susceptibility to infection, infectiousness, and other epidemiological characteristics, all of which can influence the spread of infectious diseases. Understanding these differences is important for characterizing the distribution of infection burden over an epidemic and identifying the potential for an initial infection to initiate a major outbreak. These two quantities concern different aspects of epidemic dynamics: the former describes cumulative transmission over the course of an epidemic, whereas the latter concerns the probability of sustained transmission during its early stages. Whether the age groups that experience the greatest cumulative infection burden are also those most likely to trigger a major outbreak is therefore an important question in age-structured epidemic modeling.

Empirical studies of social contacts provide an important basis for incorporating realistic mixing patterns into epidemic models. The POLYMOD study, conducted across eight European countries, systematically documented individuals' social contacts and demonstrated substantial age-assortative mixing, particularly among schoolchildren and young adults (\cite{Mossong2008}). Subsequent studies have used contact surveys and demographic information to construct age-specific contact matrices for different countries and populations. These studies highlight both the value of contact data for epidemic modeling and the need to account for differences in social contact patterns. The epidemic models incorporating empirical contact matrices can be used to study the influence of age-specific mixing pattern on epidemic outcomes.

Deterministic compartmental models, including age-structured susceptible–exposed–infectious–recovered (SEIR) models, are widely used to investigate epidemic dynamics and estimate the cumulative burden of infection. In addition to describing the temporal evolution of an epidemic, these models can yield final size relations that characterize the proportion of each population group infected over the course of an epidemic. Several studies have investigated general final-size relations for epidemic models \cite{Brauer2005,Brauer2008,Brauer2008book,Brauer2012}. Brauer extended the analysis to general age-of-infection models (\cite{Brauer2008ageandfinalsize,Brauer2009age}). More related work has examined heterogeneous mixing and more general epidemic structures. These results provide a theoretical foundation for estimating the group-specific final attack rate, and further evaluating the long-term infection burden in different age groups of the population. When the number of infected individuals is small, stochastic effects can determine whether transmission dies out or develops into a major epidemic. Continuous-time Markov chain (CTMC) models and the early-stage branching-process approximations provide a framework for quantifying this uncertainty. General stochastic epidemic modeling and approaches for approximating probabilities of a minor/major epidemic outbreak have been investigated in \cite{Allen2010,Allen2017}. Related work \cite{Bai2020} investigated final epidemic sizes in deterministic models and the probabilities of minor outbreaks in corresponding CTMC models, illustrating that epidemic outcomes can differ even when models share the same basic reproduction number $\mathcal{R}_0$. In age-structured settings, CTMC models and the approach of branching process approximation can characterize how an initial infectious individual from different age groups affects the probability of initiating a major outbreak. This probability quantifies the potential for an initial infection to trigger a major outbreak and can provide useful information for designing public health policy.

Despite the extensive literature on empirical contact patterns, deterministic final size relation calculations, and stochastic epidemic modeling with estimations of epidemic outbreak probabilities, an important question remains: Do the age groups with the greatest cumulative infection burden also have the highest probability of triggering a major epidemic outbreak? In this study, we combine an age-structured SEIR model with a CTMC-based branching-process analysis to examine the relationship between group-specific final attack rates and early-stage outbreak-trigger probabilities. Using empirical age-specific contact data for Guangzhou, Foshan, and Zhuhai, China, we investigate this relationship for two illustrative disease scenarios: COVID-19, and hand, foot, and mouth disease (HFMD). We also compare the age-specific rankings of the two measures within each city, and examine how the agreement of the rankings varies across two disease scenarios and different contact patterns in three cities. Implications for public health managements are further discussed based on case studies on two diseases. Finally, we discuss the limitations of the present study and outline directions for future research.

\medspace

\section{Data}\label{section2}
\subsection{Demographic data}
The baseline population demographic profiles and age-stratified cohort sizes for the simulations of transmissions of infectious diseases are typically sourced from the official databases of the United Nations Population Division (\citep{UN_wpp_2024}). To capture highly localized urbanization patterns and demographic distributions at the sub-national municipal level, the population metrics for individual Chinese cities can also be derived directly from the localized registries of the Seventh National Population Census of the People's Republic of China (\citep{NBS_seventh_census_2021}). Utilizing the municipal census baseline ensures that the age-specific denominator weights embedded in the contact matrices reflect the precise, real-world population structures of the target urban districts, thereby minimizing demographic skewing and statistical bias in our mathematical models.

\subsection{Social contact data and the contact matrix}
Cross-sectional social contact surveys are widely used to quantify age-specific mixing patterns in a population~\cite{Mossong2008}. In such surveys, participants report their own age and other characteristics, together with the number of contacts they made in various locations during a typical day. The population is partitioned into $n$ subgroups, and $C_{ij}$ denotes the mean number of contacts that a
randomly selected individual in subgroup $i$ makes with individuals in subgroup $j$, $i,j=1,\dots,n$, estimated from the survey. The subgroup sizes
$\mathbf{N}=(N_1,\dots,N_n)$ are obtained from census data. To satisfy the balanced nature of contacts in the population (\cite{Wallinga2006,Bai2021,Brauer2008,Jacquez1988}), the updated balanced mean contact number is
\begin{equation*}
    \tilde{C}_{ij} = \dfrac{1}{2N_i}(N_i C_{ij}+N_j C_{ji}).
\end{equation*}
$\Tilde{\mathbb{C}}_{n\times n}$ is denoted as the mean contact number matrix with elements $\Tilde{C}_{ij}$. The contact rates $\beta_i$, defined as the mean number of contacts made by an individual in subgroup $i$ per day, are estimated by 
\begin{equation*}
    \beta_i = \sum_{j=1}^n \Tilde{C}_{ij}, \quad i=1,\cdots,n. 
\end{equation*}
Furthermore, we define $p_{ij}$ as the proportion of contacts made by an individual in subgroup $i$ are with individuals in subgroup $j$.
\begin{equation*}
    p_{ij} = \dfrac{\tilde{C}_{ij}}{\beta_i}.
\end{equation*}
Consequently, the relationship between the contact matrix $\mathbb{P}$ and the mean contact number matrix $\Tilde{\mathbb{C}}$ can be established,
\begin{equation}\label{generateRandomMatrix}
\Tilde{\mathbb{C}} = \operatorname{diag}(\beta_1,\beta_2,\cdots,\beta_n)\,\mathbb{P}.  
\end{equation}
Where $\operatorname{diag}(\beta_1,\beta_2,\cdots,\beta_n)$ is the invertible $n\times n$ matrix with diagonal values $\beta_1,\beta_2,\cdots,\beta_n$.

\medspace
\subsection{Other age-dependent epidemiological effects}
Age-related heterogeneities are fundamental to understanding and modeling infectious disease dynamics \citep{Anderson1985}. In particular, age-specific variations in three key epidemiological parameters incubation period, infectious period, and transmission probability have a major influence on population-level outbreak trajectories. Pathogens such as influenza \citep{Greer2010}, COVID-19 \citep{Davies2020}, chickenpox \citep{Ferguson1996}, and hand-foot-and-mouth disease (HFMD) demonstrate how these epidemiological heterogeneities shape transmission dynamics and dictate targeted containment strategies.

First, age-dependent incubation periods govern the temporal lag between initial exposure and clinical onset, determining the intrinsic epidemic expansion mechanism. For instance, short incubation periods in pediatric HFMD cases drive rapid symptom emergence and compressed transmission cycles in early childhood settings, requiring models to account for age-varying latent phases to accurately forecast peak timing and early intervention windows. Second, age-specific variations in the duration of infectiousness shape overall outbreak persistence and community infection risk. Children with HFMD frequently exhibit prolonged shedding post-symptom resolution, extending exposure risk across household and school contacts, whereas older COVID-19 patients exhibit distinct viral dynamics coupled with elevated clinical severity. Third, transmission probability reflects the combined effects of age-stratified susceptibility and infectivity. In younger aged children, high baseline susceptibility paired with dense physical interactions in daycare settings elevates effective transmission per contact. Incorporating these age-heterogeneous epidemiological parameters allows epidemic models to make more accurate predictions, providing a mechanistic foundation for designing and evaluating age-targeted public health measures.

\medspace

\section{Mathematical models and some results}\label{sect3}
\subsection{The deterministic epidemic model and the final size relation}
We first formulate the general SEIR (susceptible-exposed-infected-recovered) epidemic ODE model. The population is stratified by age, which constitutes different subgroups. In each subgroup, the population is divided into the susceptible individual group $S_i$, the exposed individual group $E_i$, the infected individual group $I_i$, and the recovered individual group $R_i$ (for $i=1,\cdots,n$). The population size of the subgroup $i$ is $N_i = \alpha_i N = S_i + E_i + I_i + R_i$. $p_{ij}$ ($i,j=1,\cdots, n$) is an element of the contact matrix $\mathbb{P}$ and represents the probability of contacts made by a member of the group $i$ with a member of the group $j$ (\cite{Bai2020,Bai2021,Brauer2008}). We assume that the average incubation period for each subgroup is $\displaystyle\frac{1}{\delta_i}$, and the average infectious period in each subgroup is $\displaystyle\frac{1}{\gamma_i}$ ($i=1,\cdots,n$). The transmission probability specified for the disease is $\epsilon_{ij} \ll 1$,  where it represents the probability of transmission given a contact event between a susceptible individual in age group $i$ and an infectious individual in age group $j$. It can be interpreted as the product of the baseline scalar associated with the disease, the susceptibility of age group $i$ and the infectivity of age group $j$. The epidemic model with the contact matrix $\mathbb{P}$ can be expressed as
\begin{equation}\label{mainmodel}
	\begin{cases}
	S_{i}^{\prime} &= - S_{i}\beta_i\sum\limits_{j=1}^{n}\epsilon_{ij}p_{ij}\displaystyle\frac{I_j}{N_j} \\
	E_{i}^{\prime} &= S_{i}\beta_i\sum\limits_{j=1}^{n}\epsilon_{ij}p_{ij}\displaystyle\frac{I_j}{N_j} - \delta_i E_i \\
    I_i^{\prime} &= \delta_i E_i - \gamma_i I_i  \\
	R_{i}^{\prime} &= \gamma_i I_{i},
\end{cases}
\end{equation}
for $i = 1, \cdots, n$. $\epsilon_{ij} = q\times u_i \times v_j$, where $u_i$ represents the relative susceptibility of individuals in group $i$, $v_j$ represents the relative infectivity of individuals in group $j$, and $q$ is an undetermined scaling parameter. It is worth mentioning that some requirements must be satisfied for the contact matrix $\mathbb{P}$,
\begin{equation}\label{ConstraintsP}
    \begin{aligned}
       & p_{ij} \geq 0, \quad \sum_{j=1}^n p_{ij} = 1, \\
       & N_i\beta_i p_{ij} = N_j\beta_j p_{ji}.
    \end{aligned}
\end{equation}
The second requirement is called the balance equation. The next generation matrix $FV^{-1}$ for the model \eqref{mainmodel} can be obtained as (\cite{vandenDriessche2002,Diekmann2009}),
The basic reproduction number $\mathcal{R}_0 = \rho(FV^{-1})$ is the spectral radius of the next generation matrix (\cite{Diekmann2009}). The derivation of $\mathcal{R}_0$ is presented in Appendix~\ref{app:R0}. The explicit formula for $\mathcal{R}_0$ can only be obtained for lower-dimensional matrices $(n \leq 3)$ and certain special types of contact patterns. The final size relation for each subgroup is
\begin{equation}\label{final_size_relation}
\log\displaystyle\frac{N_i}{S_i(\infty)} = \beta_i\sum_{j=1}^n \dfrac{\epsilon_{ij}p_{ij}}{\gamma_j}\displaystyle\frac{N_j-S_j(\infty)}{N_j}.
\end{equation}
The long-term risk of infection in each subgroup during the epidemic process can be calculated as
\begin{equation*}
\pi_{i} = \displaystyle\frac{N_i-S_i(\infty)}{N_i}, \quad i=1,\cdots,n.    
\end{equation*}
The standard derivation of the final size relation can be found in Appendix~\ref{app:derivationsDet} (See \cite{Brauer2012} for more discussions).
\medspace

\subsection{The stochastic epidemic model and the probability of a major outbreak}\label{sect4}
The random variable $(S_{i,t},E_{i,t},I_{i,t},R_{i,t})$ with $(i=1,\cdots,n)$ is the state vector that represents the population sizes of the susceptible class, the exposed class, the infectious class, and the removed class (recovered) in age group $i$ at time $t$, respectively. For any time $t \geq 0$ and $i=1,\cdots, n$, $S_{i,t} \geq 0$, $E_{i,t}\geq 0$, $I_{i,t}\geq 0$, and $R_{i,t} \geq 0$, the random variables are upper bounded by the population sizes of the subgroups. The stochastic process can be described as
\begin{equation}\label{stochasticSIR}
	\begin{cases}
		\mathbb{P}((S_{i,t+\Delta t},E_{i,t+\Delta t},I_{i,t+\Delta t},R_{i,t+\Delta t})-(S_{i,t},E_{i,t},I_{i,t},R_{i,t})=(-1,1,0,0))&=s_{i}\beta_i\sum\limits_{j=1}^{n}\epsilon_{ij}p_{ij}\displaystyle\frac{i_j}{N_j}\Delta t + o(\Delta t), \\
        \mathbb{P}((S_{i,t+\Delta t},E_{i,t+\Delta t},I_{i,t+\Delta t},R_{i,t+\Delta t})-(S_{i,t},E_{i,t},I_{i,t},R_{i,t})=(0,-1,1,0))&=\delta_i e_i\Delta t + o(\Delta t),\\
		\mathbb{P}((S_{i,t+\Delta t},E_{i,t+\Delta t},I_{i,t+\Delta t},R_{i,t+\Delta t})-(S_{i,t},E_{i,t},I_{i,t},R_{i,t})=(0,0,-1,1))&=\gamma_i i_i\Delta t + o(\Delta t),
	\end{cases}
\end{equation}
with the complementary probability
\begin{equation*}
\begin{aligned}
	\mathbb{P}((S_{i,t+\Delta t},E_{i,t+\Delta t},I_{i,t+\Delta t},R_{i,t+\Delta t})&-(S_{i,t},E_{i,t},I_{i,t},R_{i,t})=(0,0,0,0))=\\
	&1-\sum\limits_{i=1}^{n}\sum\limits_{j=1}^{n}s_{i}\beta_i\epsilon_{ij}p_{ij}\displaystyle\frac{i_j}{N_j}\Delta t-\sum\limits_{i=1}^{n}\delta_i e_i\Delta t-\sum\limits_{i=1}^{n}\gamma_i i_i\Delta t + o(\Delta t).\end{aligned}
\end{equation*}
$(s_i,e_i,i_i,r_i)$ represents the state value of the random variable $(S_{i,t},E_{i,t},I_{i,t},R_{i,t})$. For each Markov jump process, there is an exponentially distributed time between every jump (\citep{Allen2010, Allen2017, Greenwood2009}). The complete list of all infinitesimal transition probabilities for model \eqref{stochasticSIR} is defined in Table \eqref{tab1}.

\medskip

\begin{table}[htp]
\caption{Discrete events of the CTMC model \eqref{stochasticSIR} and the infinitesimal transition  probabilities }\label{tab1}
\begin{center}
\begin{tabular}{|p{5cm}| p{2.93cm}| p{6cm}|}
\hline
\toprule
\textbf{Event} & \textbf{Transition} & \textbf{Probability} \\
\hline
\hline

Incident of new infection in age group $1$ & $S_1 \rightarrow S_1 - 1$\newline$E_1 \rightarrow E_1 + 1$ & $s_{1}\beta_1\sum\limits_{j=1}^{n}\epsilon_{1j}p_{1j}\frac{i_j}{N_j}\Delta t+o(\Delta t)$\\
\midrule
Transition from exposed stage to infectious stage in group $1$ & $E_1\rightarrow E_1-1$\newline $I_1\rightarrow I_1+1$ & $\delta_1 e_1\Delta t+o(\Delta t)$
\\
\midrule
New recovery in age group $1$ & $I_1\rightarrow I_1-1$ \newline $R_1\rightarrow R_1+1$ & $\gamma_1 i_1\Delta t+o(\Delta t)$
\\
\midrule
$\vdots$ & $\vdots$ & $\vdots$\\
\midrule
Incident of new infection in age group $n$ &$S_n \rightarrow S_n - 1$\newline$E_n \rightarrow E_n + 1$ & $s_{n}\beta_n\sum\limits_{j=1}^{n}\epsilon_{nj}p_{nj}\frac{i_j}{N_j}\Delta t+o(\Delta t)$\\
\midrule
Transition from exposed stage to infectious stage in group $n$ & $E_n\rightarrow E_n-1$\newline $I_n\rightarrow I_n+1$ & $\delta_n e_n\Delta t+o(\Delta t)$
\\
\midrule
New recovery in age group $n$ & $I_n\rightarrow I_n-1$ \newline $R_n\rightarrow R_n+1$ & $\gamma_n i_n\Delta t+o(\Delta t)$\\
\hline
\bottomrule
\end{tabular}
\end{center}
\label{Table1}
\end{table}

Probabilistic coupling yields the result that during the early stages of an epidemic, the process of an outbreak can be approximated by a branching process. Therefore, we apply the theory of the multi-type branching process (\cite{Allen2010,Allen2017,Athreya1972}) to investigate the behavior near the disease-free equilibrium,

$$(S_1,E_1,I_1,R_1,S_2,E_2,I_2,R_2,\cdots, S_n,E_n,I_n,R_n) = (N_1,0,0,0,N_2,0,0,0,\cdots,N_n,0,0,0).$$ 

The probability generating functions (pgfs) and the backward Kolmogorov
differential equation are subsequently used to obtain the estimate of the probability of a minor (major) outbreak (\cite{Allen2010,Allen2017}). We study $2n$ infection-related variables, $E_i$ and $I_i$ (where $i=1,\cdots,n$), through the formulation of probability-generating functions. The domain of each pgf is $u \in [0,1]^{2n}$, where $u=(u_{11},u_{12},u_{21},u_{22}\cdots,u_{n1},u_{n2})$. The pgf represents the number of ‘offspring’ generated by infectious individuals. In order to apply pgfs from the Branching process approximation to the Markov chain process, we assume that the initial number of infectious individuals is small and that the ‘birth’ of new infectious individuals occurs independently of each other (see \cite{Allen2010,Allen2017}). For each pgf, it is assumed that there is only one initial exposed or infectious individual, \ie, for $f_{11}(u)$, $E_1(0)=1$, $E_i(0)=0$ for $i \neq 1$ and $I_{i}(0) = 0$ for $i=1,\cdots,n$; for $f_{12}(u)$, $I_1(0)=1$, $I_i(0)=0$ for $i \neq 1$ and $E_{i}(0) = 0$ for $i=1,\cdots,n$. Each of the $2n$ pgfs is defined in the birth-and-death process. For the continuous-time Markov chain model \eqref{stochasticSIR}, the pgfs are expressed in the following formula:
\begin{equation}\label{pgfs}
\begin{cases}
f_{11}(u) &= \displaystyle\frac{\delta_1 u_{12}}{\delta_{1}} = u_{12} \\
f_{12}(u) &= \displaystyle\frac{\displaystyle\sum\limits_{i=1}^{n}\beta_i \epsilon_{i1} p_{i1}\displaystyle\frac{N_i}{N_1}u_{12} u_{i1} + \gamma_1}{\sum\limits_{i=1}^{n}\beta_i \epsilon_{i1} p_{i1}\displaystyle\frac{N_i}{N_1} + \gamma_1}, \\
\cdots & \\
f_{n1}(u) &= \displaystyle\frac{\delta_n u_{n2}}{\delta_{n}} = u_{n2} \\
f_{n2}(u) &= \displaystyle\frac{\displaystyle\sum\limits_{i=1}^{n}\beta_i\epsilon_{in} p_{in}\displaystyle\frac{N_i}{N_n}u_{n2} u_{i1} + \gamma_n}{\displaystyle\sum\limits_{i=1}^{n}\beta_i\epsilon_{in} p_{in}\displaystyle\frac{N_i}{N_n} + \gamma_n}.
\end{cases}    
\end{equation}
It is well known that there is at least one fixed point of the pgfs in the domain $u \in [0,1]^{2n}$, namely $(1, 1, \cdots, 1)$. However, there exists at least one more fixed point when the basic reproduction number $\mathcal{R}_0 > 1$. It follows from the multi-type branching process theory that the minimal non-negative fixed point of the pgfs $(q_{11}, q_{12} \cdots, q_{n1}, q_{n2})$ is the asymptotic estimate for the probability of a minor outbreak given either one latent individual or one infectious individual (\cite{Athreya1972,Allen2010,Allen2017,Bai2020,Bai2018}). The derivation and related results are presented in Appendix~\ref{app:derivationSto} (More detailed discussions can be found in \cite{Allen2010,Allen2017}). 
From \eqref{pgfs}, since the death rates in exposed stages for all age groups are not considered, we have the following,
\begin{equation*}
    q_{i1} = q_{i2}, \quad i=1,\cdots,n.
\end{equation*}
Therefore, it is convenient to denote $q_{i}:=q_{i1}=q_{i2}$ for $i=1,\cdots,n$. Thus, for the random variable $X(t)$ corresponding to the $n$ infection-related states corresponding to $n$ age groups, and $X(0)=(0,0,\cdots,0,1,0,\cdots,0)$ with $1$, the non-zero element represents the initial condition of an exposed or infectious individual in the age group $i$, the conditional probability of a minor outbreak of an epidemic is defined as
\begin{equation}\label{definition1}
\lim_{t\rightarrow\infty}\mathbb{P}(X(t)=0|X(0)) = q_{i}.
\end{equation}
If we further consider the nature of the balance relation described in \eqref{ConstraintsP}, the system of implicit equations that represent the asymptotical probability of a minor outbreak can be written as
\begin{equation}\label{simpliedpgfs}
\begin{cases}
    q_1 &= \displaystyle\frac{\displaystyle\sum\limits_{i=1}^{n}\beta_1 \epsilon_{i1} p_{1i}q_1 q_i + \gamma_1}{\sum\limits_{i=1}^{n}\beta_1 \epsilon_{i1} p_{1i} + \gamma_1}, \\
\cdots & \\
q_n &= \displaystyle\frac{\displaystyle\sum\limits_{i=1}^{n}\beta_n\epsilon_{in} p_{ni}q_n q_i + \gamma_n}{\displaystyle\sum\limits_{i=1}^{n}\beta_n\epsilon_{in} p_{ni} + \gamma_n}.
\end{cases}    
\end{equation}

\medspace

\section{Applications}
We demonstrate our methods using empirical social contact data from Guangzhou, Foshan, and Zhuhai—three key cities within China's highly integrated Guangdong-Hong Kong-Macao Greater Bay Area (GBA). These cities provide a representative cross-section of the region's diverse urban densities, industrial hubs, and cross-border transit dynamics. The age-stratified population sizes for three cities are presented in Table \eqref{tab:population_sizes}. The contact survey data were originally collected and analyzed by Huang et al. \cite{Huang2020}, and the corresponding age-stratified contact matrices were generated using the \texttt{socialmixr} package \cite{Funk2018} in R. The generated contact matrices for each city are incorporated into deterministic and stochastic epidemic models discussed in Section \eqref{sect3}. We apply the deterministic modeling framework to accurately predict long-term epidemiological outcomes, including final sizes and age-stratified attack rates in all age groups. The stochastic modeling framework is used to assess initial transmission risks and determine which age groups are most vulnerable to initiating an epidemic outbreak.

ChatGPT (OpenAI) and Claude (Anthropic) were used to assist in developing, debugging, and refining the R code used for numerical simulations and data visualization. The authors reviewed and verified the computational procedures and numerical results and take full responsibility for their accuracy and reproducibility.

\medspace

\begin{table}[htbp]
\centering
\caption{Age-group population sizes for Guangzhou, Foshan, and Zhuhai (\cite{NBS_seventh_census_2021})}
\label{tab:population_sizes}
\begin{tabular}{lrrr}
\toprule
Age group & Guangzhou & Foshan & Zhuhai \\
\midrule
{[}0,5)    &    915{,}000 &   512{,}000 &  142{,}000 \\
{[}5,10)   &    860{,}189 &   480{,}000 &  125{,}000 \\
{[}10,15)  &    815{,}000 &   444{,}000 &  120{,}000 \\
{[}15,20)  &  1{,}050{,}000 &   490{,}000 &  138{,}000 \\
{[}20,25)  &  1{,}300{,}000 &   600{,}000 &  160{,}000 \\
{[}25,30)  &  1{,}820{,}000 &   890{,}000 &  245{,}000 \\
{[}30,35)  &  2{,}200{,}000 & 1{,}100{,}000 &  290{,}000 \\
{[}35,40)  &  1{,}920{,}000 &   960{,}000 &  240{,}000 \\
{[}40,45)  &  1{,}600{,}000 &   800{,}000 &  240{,}000 \\
{[}45,50)  &  1{,}490{,}000 &   790{,}000 &  210{,}000 \\
{[}50,55)  &  1{,}410{,}000 &   720{,}000 &  184{,}000 \\
{[}55,60)  &  1{,}136{,}083 &   552{,}000 &  180{,}000 \\
{[}60,65)  &    700{,}000 &   400{,}000 &  100{,}000 \\
{[}65,70)  &    600{,}333 &   320{,}000 &   80{,}000 \\
70+        &    860{,}000 &   442{,}000 &   86{,}000 \\
\midrule
Total      & 18{,}676{,}605 & 9{,}500{,}000 & 2{,}540{,}000 \\
\bottomrule
\end{tabular}
\end{table}

\medskip

\subsection{Contact patterns in three cities of Greater bay area in China}

\begin{figure}[htbp]
    \centering
    \includegraphics[width=\linewidth]{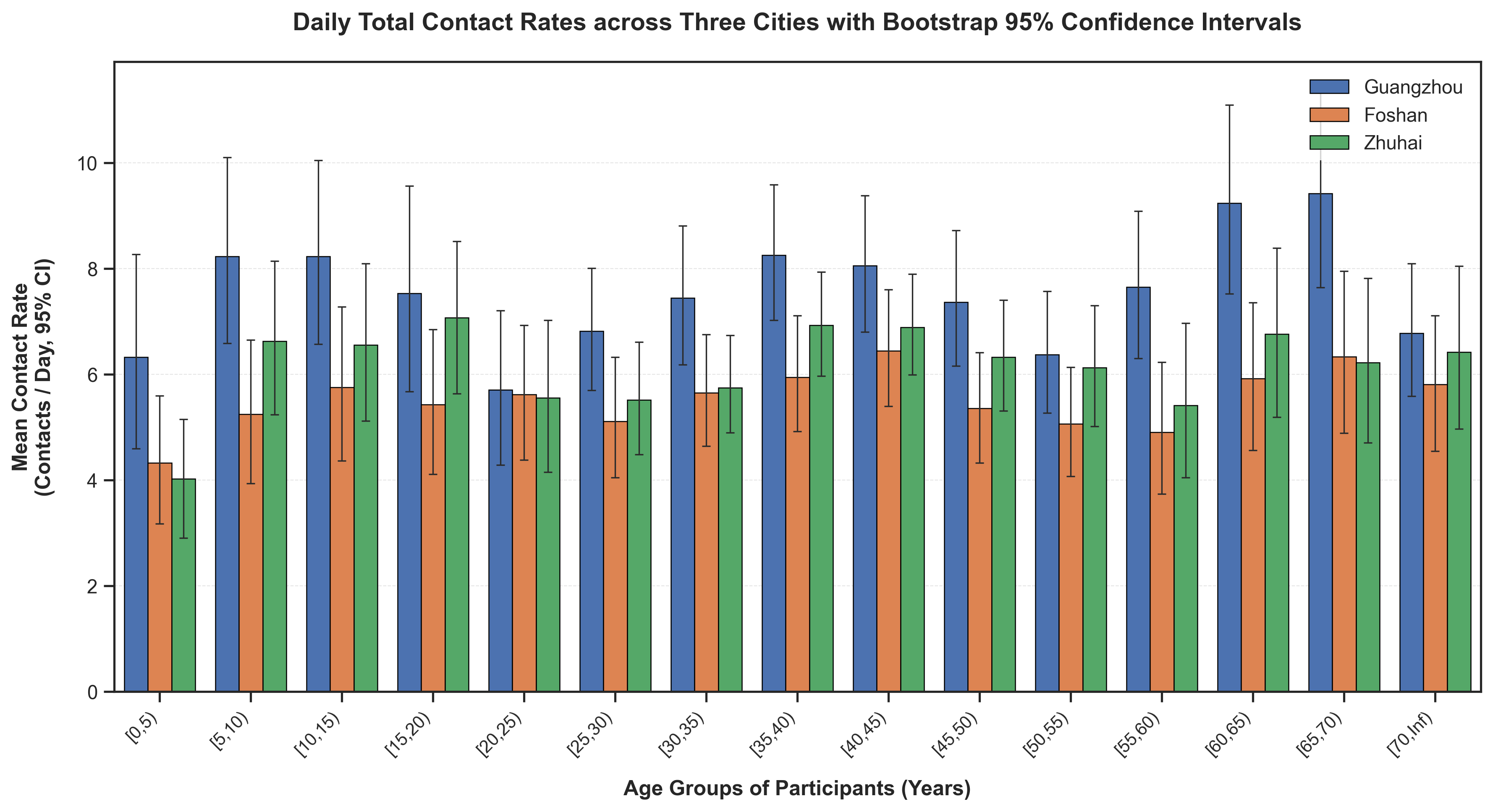}
    \caption{Averaged daily total contact rates across age groups in Guangzhou, Foshan, and Zhuhai. Bars represent the mean daily contact numbers calculated from recorded social contact data. Vertical error bars denote the $95\%$ confidence intervals (CIs) derived from bootstrap resampling.}
    \label{fig:comparison_contact_rates}
\end{figure}

We recorded the averaged contact numbers in each age group of three surveyed municipalities, and applied the bootstrap resampling method to calculate the $95\%$ confidence intervals of contact numbers. The results are visualized in Figure \eqref{fig:comparison_contact_rates}. It is worth emphasizing that the paper \cite{Huang2020} reported a significantly higher average number of daily contacts because they explicitly incorporated \textbf{supplementary professional contacts (SPCs)}. In this study, we focus strictly on discrete, individually recorded conversational and physical contact events extracted from the primary participant diaries. Importantly, this methodological divergence does not alter the underlying structural insights of our descriptive analysis, nor does it affect the validity of the mathematical models built upon the contact matrices.  

Substantial heterogeneity across the three cities and among age groups can be observed in Figure \eqref{fig:comparison_contact_rates}. Guangzhou consistently exhibits the highest mean daily contact rates across all $15$ age groups, with particularly high values among older adults aged 60–70 years. In contrast, Foshan generally has the lowest contact rates, while Zhuhai exhibits intermediate values in most age groups. The differences between Foshan and Zhuhai are not uniform across age groups. For example, Zhuhai has noticeably higher contact rates than Foshan among participants aged 35–40 and 60–65 years, whereas their estimates are quite similar in the 20–25 age group. It should be noted that the estimated contact numbers exclude supplementary professional contacts (SPCs). This difference in data processing should be taken into account when interpreting the age-specific contact patterns, particularly the relatively high contact numbers observed among older age groups.

\begin{figure*}[htbp]
    \centering
    
    \begin{subfigure}[b]{0.48\textwidth}  
        \centering
        \includegraphics[width=\textwidth]{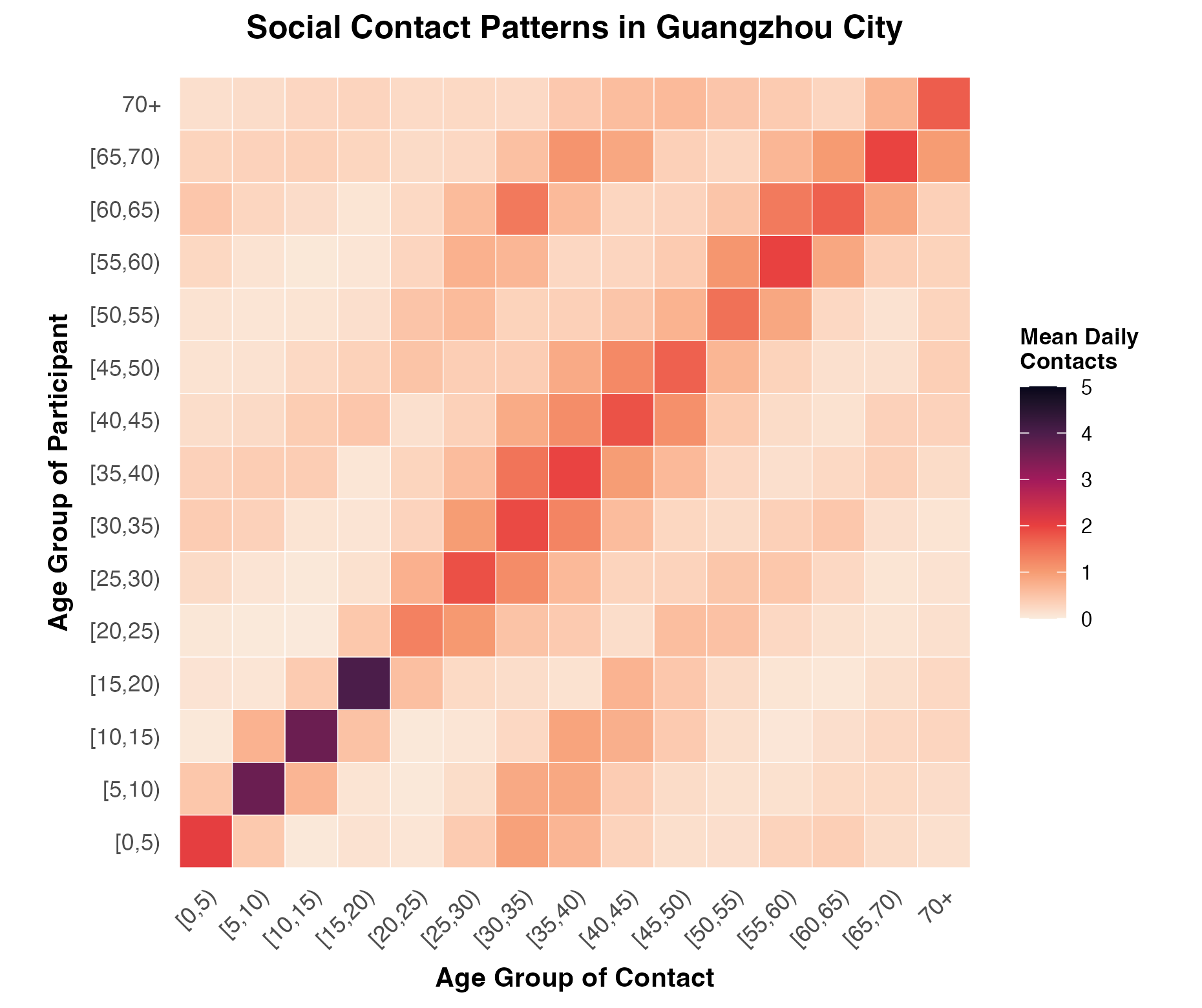}
        \caption{Guangzhou}
        \label{fig:matrix_gz}
    \end{subfigure}
    \hfill
    \begin{subfigure}[b]{0.48\textwidth}  
        \centering
        \includegraphics[width=\textwidth]{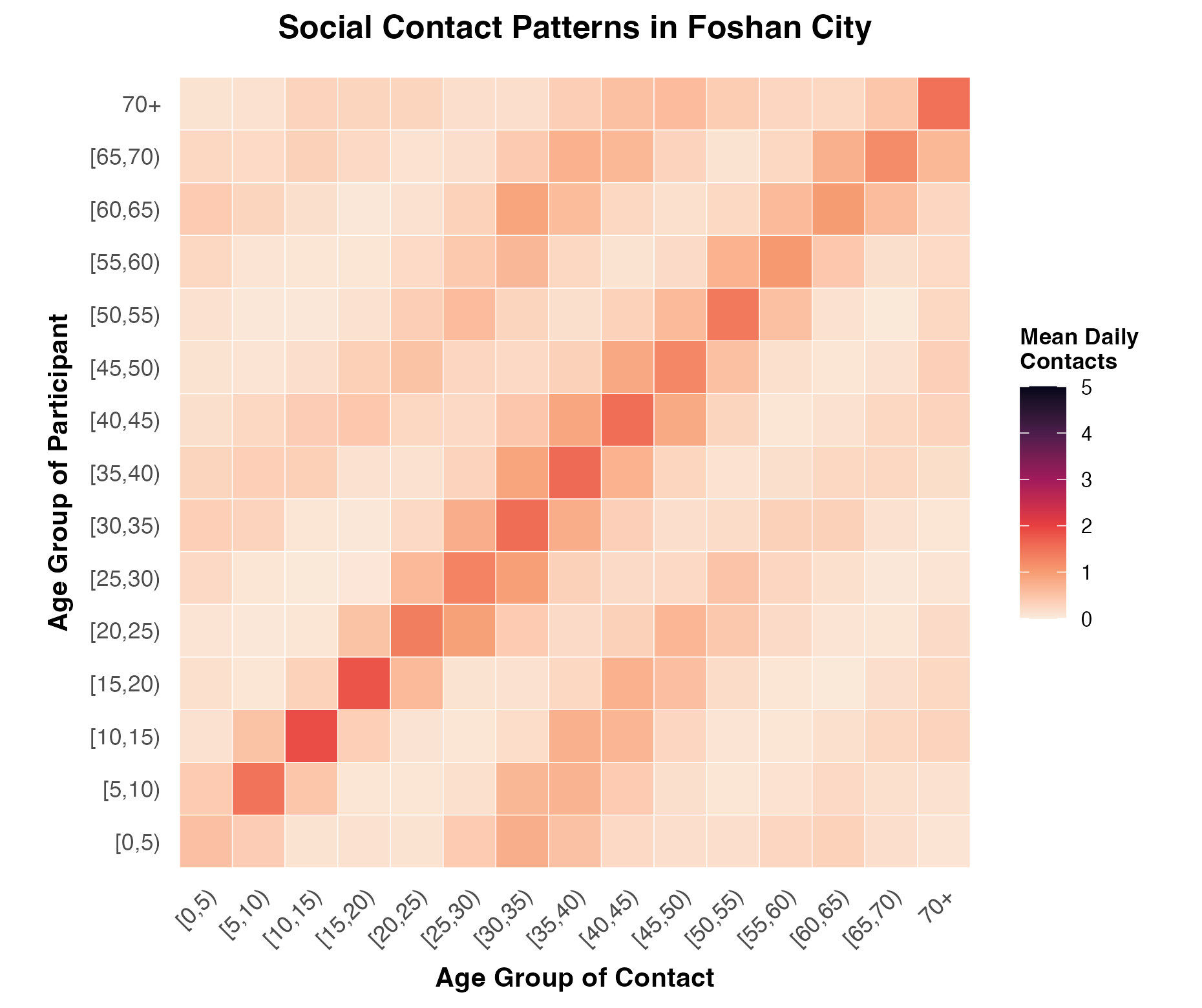}
        \caption{Foshan}
        \label{fig:matrix_fs}
    \end{subfigure}
    
    \par\vspace{1.5em}
    
    \begin{subfigure}[b]{0.48\textwidth}  
        \centering
        \includegraphics[width=\textwidth]{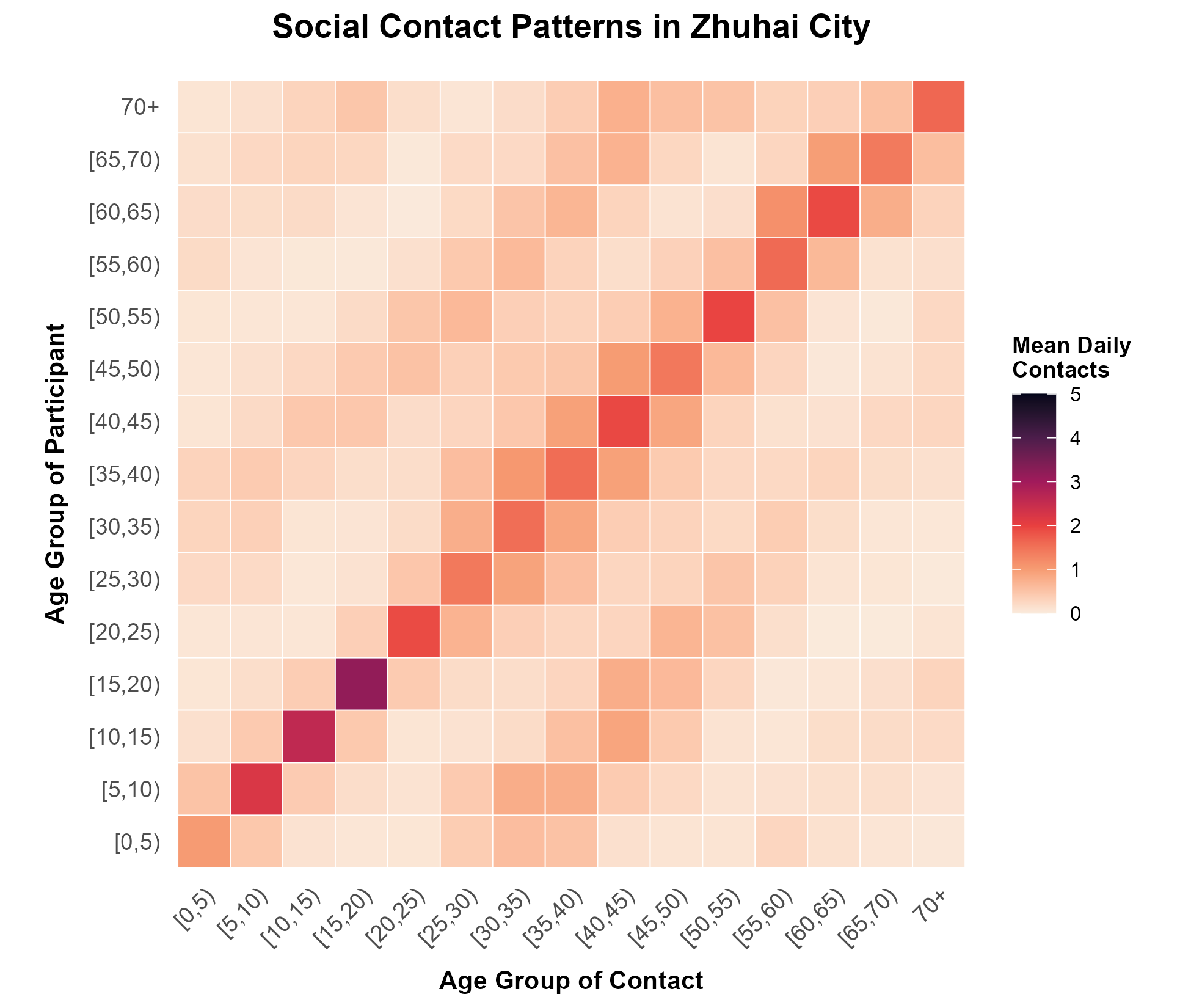}
        \caption{Zhuhai}
        \label{fig:matrix_zh}
    \end{subfigure}
    
    \caption{\textbf{Age-stratified social contact matrices. The heatmaps illustrate the mean number of daily conversational or physical contacts between participants (y-axis) and contact individuals (x-axis) across 15 stratified age cohorts (0–70+ years). All sub-figures share a standardized color scale (0 to 5 daily contacts) to facilitate direct visual comparison. Panel (a) shows intense assortative mixing in Guangzhou; Panel (b) captures low baseline mixing in Foshan; Panel (c) outlines strong cross-generational mixing patterns in Zhuhai.}}
    \label{fig:three_cities_contact_matrices}
\end{figure*}

\begin{figure}[htbp]
\centering
\begin{subfigure}[b]{0.48\textwidth}
  \centering
  \includegraphics[width=\linewidth]{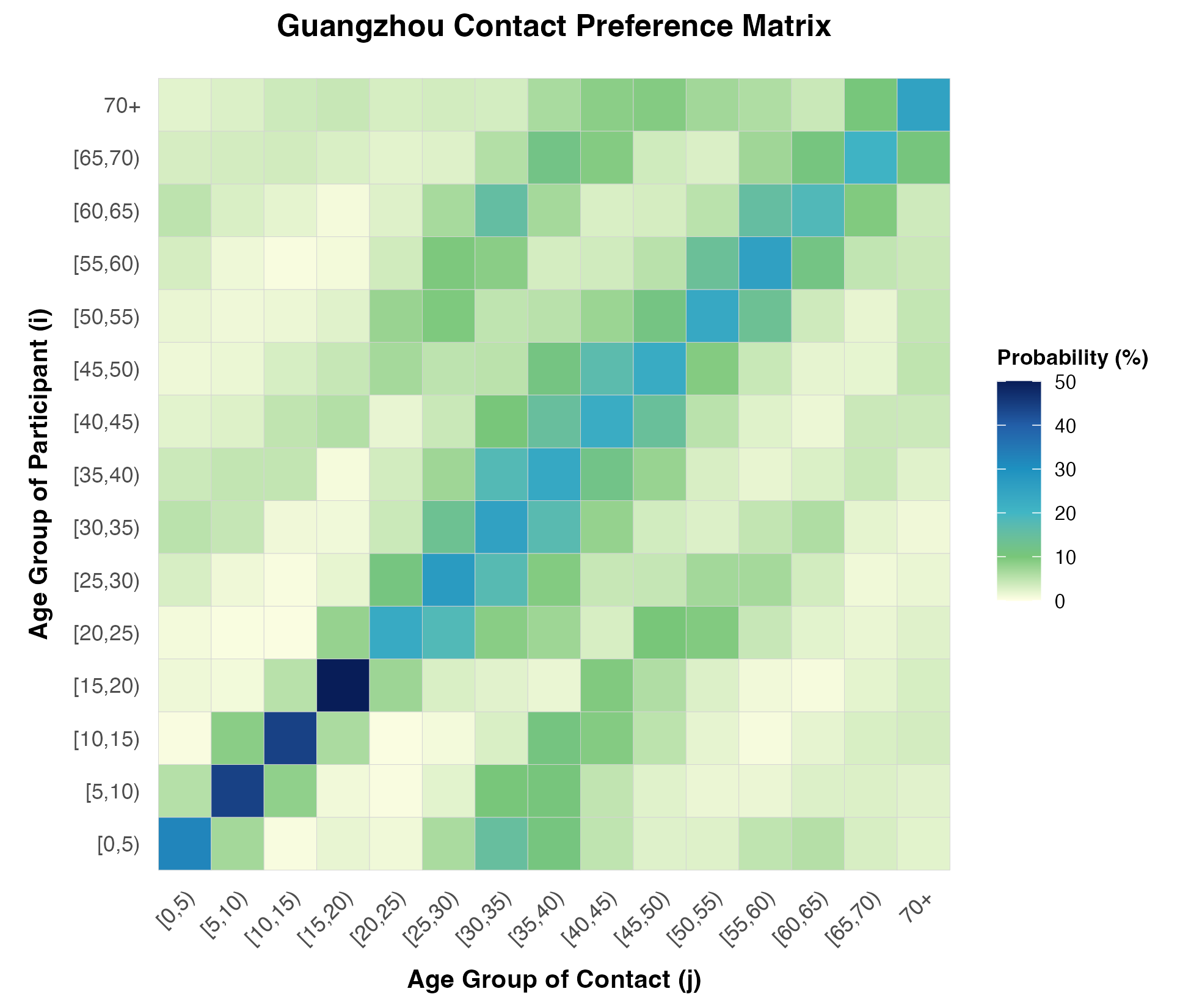} 
  \caption{Guangzhou}
  \label{fig:prob_tendency_gz}
\end{subfigure}
\hfill
\begin{subfigure}[b]{0.48\textwidth}
  \centering
  \includegraphics[width=\linewidth]{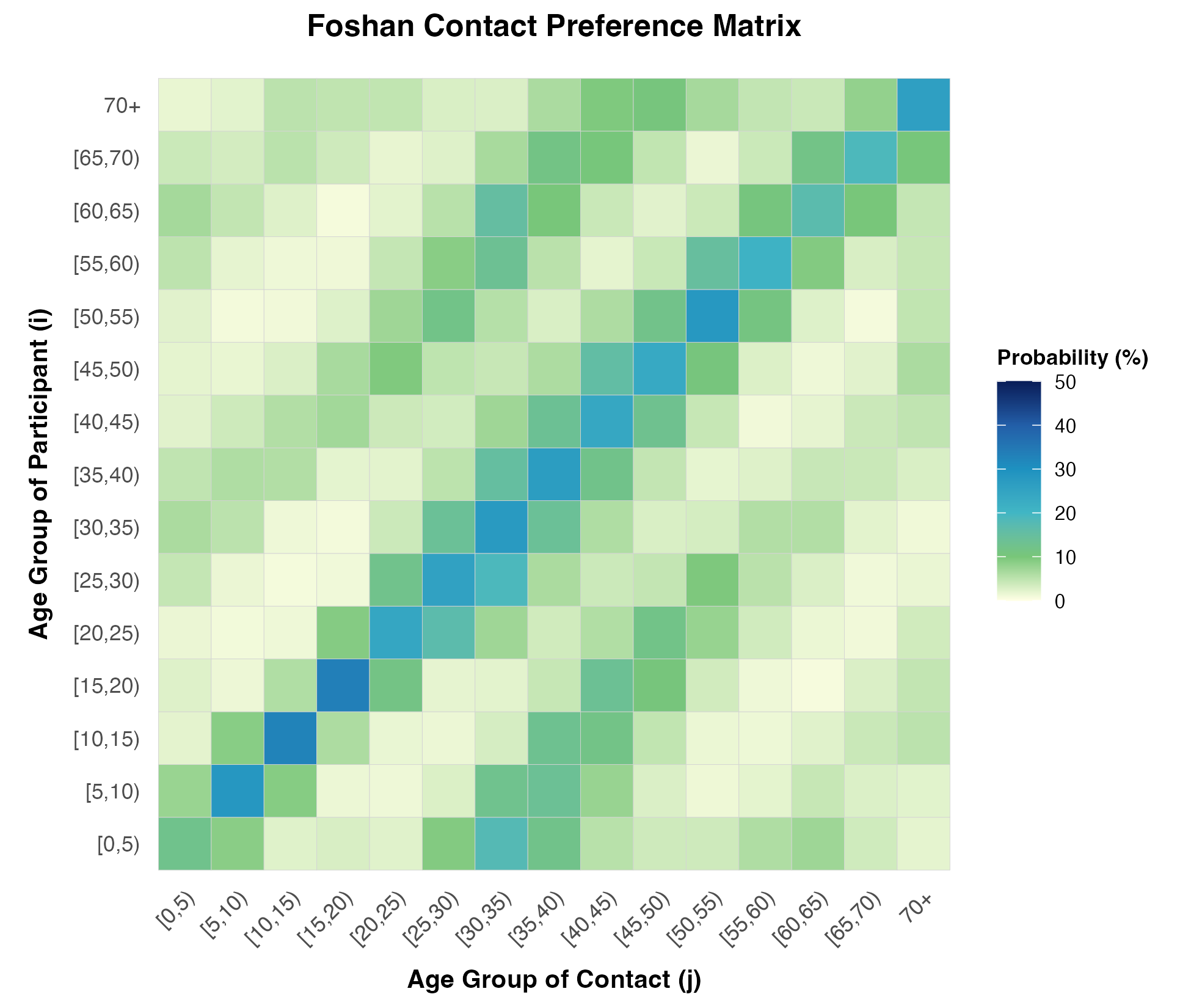}
  \caption{Foshan}
  \label{fig:prob_tendency_fs}
\end{subfigure}

\vspace{0.5cm} 

\begin{subfigure}[b]{0.50\textwidth}
  \centering
  \includegraphics[width=\linewidth]{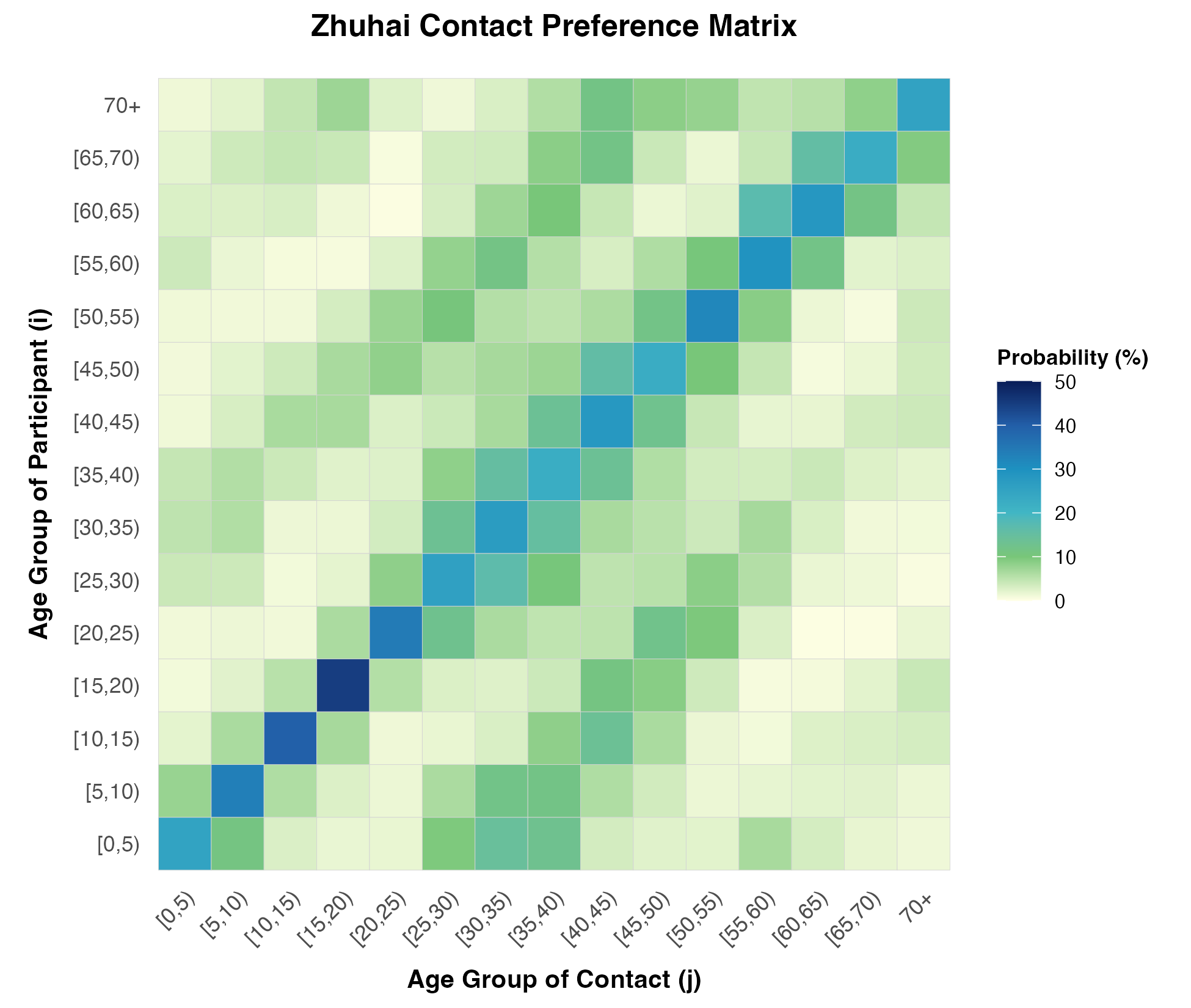}
  \caption{Zhuhai}
  \label{fig:prob_tendency_zh}
\end{subfigure}

\caption{Age-stratified social contact preference matrices ($p_{ij}$) for Guangzhou, Foshan, and Zhuhai. Elements represent the normalized probability of an individual's total daily contact capacity across 15 stratified age-cohorts, where each row sums strictly to 100\%. All panels display individual percentage color gradients scaled to their respective peak density values. Panel (a) reveals intense assortative homophily alongside a visible adult parental bridge in Guangzhou; Panel (b) captures a diffused, more uniform contact preference structure in Foshan; Panel (c) highlights strong, prominent cross-generational mixing patches representing grandparent-child contact in Zhuhai. } 
\label{fig:multi_city_probability_tendencies}
\end{figure}

Age-stratified social contact number matrices and social contact preference matrices are presented in Figures \eqref{fig:three_cities_contact_matrices} and \eqref{fig:multi_city_probability_tendencies}. The contact matrices in Figure~\eqref{fig:three_cities_contact_matrices} reveal distinct age-specific mixing patterns across the three cities. In Guangzhou, contacts are strongly concentrated along the diagonal, indicating pronounced age-assortative mixing, with additional cross-generational contacts between children and adults. Foshan exhibits a more diffuse contact pattern, with less pronounced age-specific concentration. In contrast, Zhuhai displays the most pronounced cross-generational mixing, particularly between younger children and older adults. These differences are also reflected in the corresponding contact preference matrices in Figure~\eqref{fig:multi_city_probability_tendencies}.

To demonstrate the utility of the age-structured modeling framework and the data-driven contact matrices, we present the detailed case study of the coronavirus disease $2019$ (COVID-19), caused by severe acute respiratory syndrome coronavirus $2$ (SARS-COV-$2$), and the case study of the Hand-foot-Mouth disease (HFMD) in three cities of Guangdong Province.

\medskip

\medskip

\subsection{The long-term and short-term analysis of COVID-19 epidemic}

COVID-$19$ exhibits heterogeneity across age groups. The basic reproduction number $\mathcal{R}_0$ is in the moderate range, which places the transmission process within the supercritical regime. The derived final size relation in \eqref{sect3} can be applied to estimate the final outcome of the epidemic, and the equation system in \eqref{simpliedpgfs} about probabilities of a major outbreak is able to identify the short-term outbreak risks. In this paper, we set the value of the basic reproduction number $\mathcal{R}_0=2.5$ (\cite{Li2020}). $\mathcal{R}_0$ can be calculated as the spectral radius of the next-generation matrix $K$ (\cite{Diekmann1990}), which is given by
\begin{equation*}
 K_{ij}=\frac{N_i}{N_j}\frac{\beta_i\epsilon_{ij}p_{ij}}{\gamma_j},
\qquad i,j=1,\ldots,n.
\end{equation*}
Here, $K_{ij}$ represents the expected number of new infections in age group $i$ generated by a single infectious individual in age group $j$. The basic reproduction number is therefore given by
\begin{equation*}
\mathcal R_0=\rho(K).    
\end{equation*}

To capture the biological variation in viral latency across the population, the parameter values are properly set according to previous studies. It has been shown that the incubation periods follow a U-shaped pattern with respect to age (\cite{Kong2020,Qi2021}). The infectious period exhibits a monotonically increasing relationship with age (\cite{Zhou2020}). The relative susceptibilities in age groups are varied. The children and adolescents are assigned values $0.5$, $0.55$ and $0.6$ relative to the adult baseline of $1.0$ (\cite{Viner2021}). Relative infectivity is assumed uniform for all age groups. The parameter values used in modeling are presented in Table \eqref{tab:covid_params} in Appendix~\ref{app:figuresandtables}. The scaling parameter $q$ can be extracted based on the relation $\epsilon_{ij}=q u_i v_j$ by fitting the target basic reproduction number $\mathcal{R}_0=2.5$. 

\medskip

\begin{figure}[htbp]
    \centering
    \includegraphics[width=\linewidth]{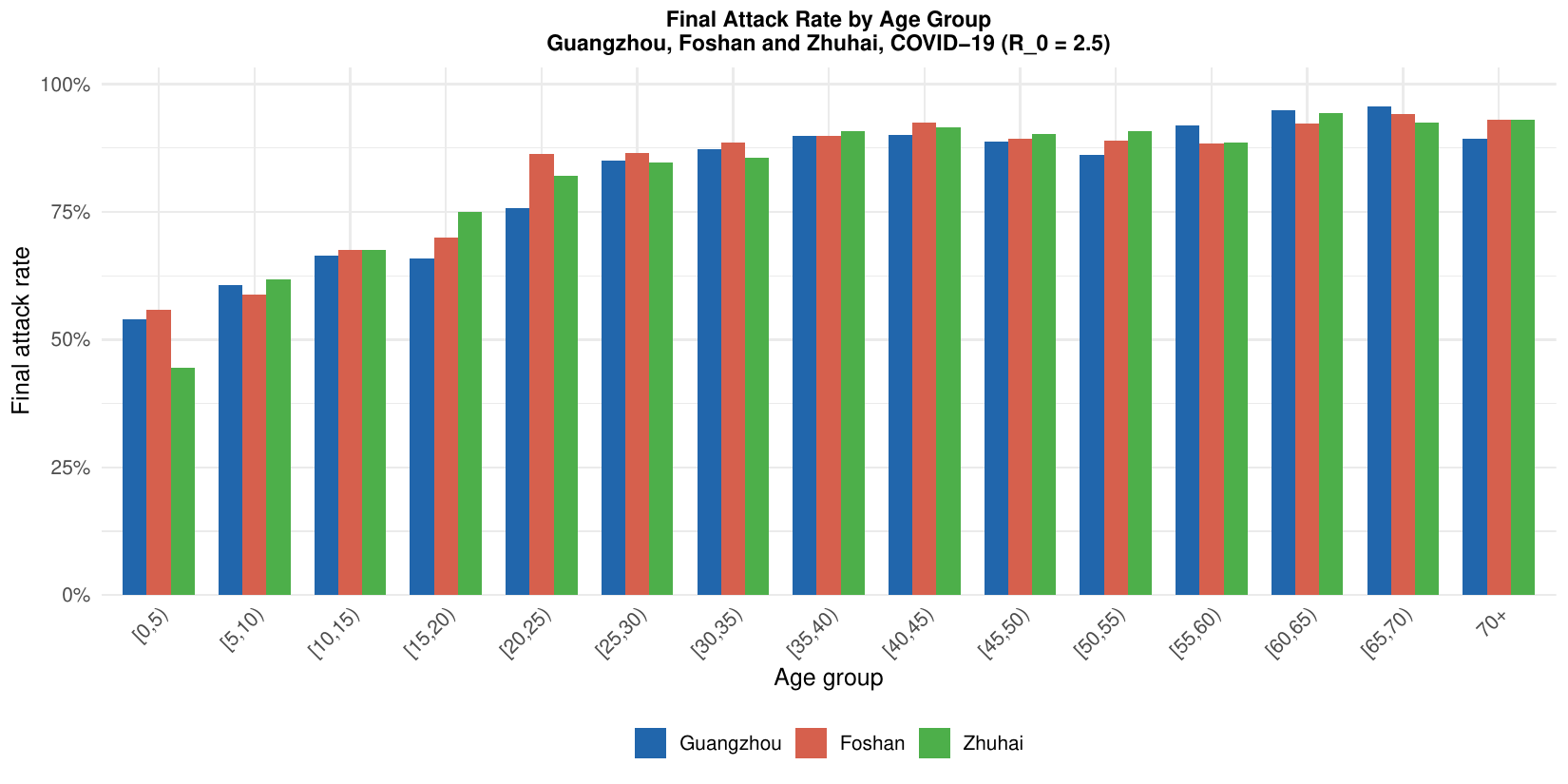}
    \caption{Final attack rate by age group across Guangzhou, Foshan and Zhuhai (COVID-19, $\mathcal{R}_0=2.5$). Bars show the deterministic age-structured attack rate for each age group, obtained from the final size relation \eqref{final_size_relation} applying the empirical contact matrices in three cities and age-stratified epidemiological parameters. Colors denote city: Guangzhou (blue), Foshan (orange), Zhuhai (green).}
    \label{fig:comparison_attack_rates}
\end{figure}

The final attack rates by age group in all three cities are illustrated in Figure \eqref{fig:comparison_attack_rates}. In all three cities, the final attack rate rises monotonically with age from the youngest group through the $[60,65)$ group, before declining slightly in the $70+$ age group. The population-weighted attack rates in the three cities are similar, with values of $82.8\%$, $84.1\%$, and $83.2\%$. For comparison, the attack rate predicted under homogeneous mixing is $89.3\%$ for all three cities (presented in Figure \eqref{fig:ar_three_cities} in Appendix~\ref{app:figuresandtables}). This gap indicates that ignoring contact heterogeneity leads to a systematic overestimation of public health burden (\cite{Bai2020}). The longer infectious periods among older adults are possibly the strongest driving factors behind the rising attack rate observed from young adulthood through the oldest age groups, while reduced susceptibility likely accounts for the relatively lower attack rates among children and adolescents. It is notable that the $[0,5)$ age group in Zhuhai suffers the lowest attack rate ($45\%$) among the three cities. This is not simply explained by the lowest overall contact rate for this age group ($4.02$), since Foshan's $[0,5)$ age group has a similar overall contact rate ($4.33$) yet suffers a markedly higher attack rate of $56\%$. The same age group in Guangzhou has a higher overall contact rate ($6.32$) but a lower attack rate than Foshan. This observation suggests that the full structure of the contact pattern, including overall contact rate, within-age-group assortativity, and the distribution of contacts across other age groups, jointly governs epidemiological outcomes and drives the observed variation in final attack rates in age groups.

\medspace

\medspace

\begin{figure}[htbp]
    \centering
    \includegraphics[width=\linewidth]{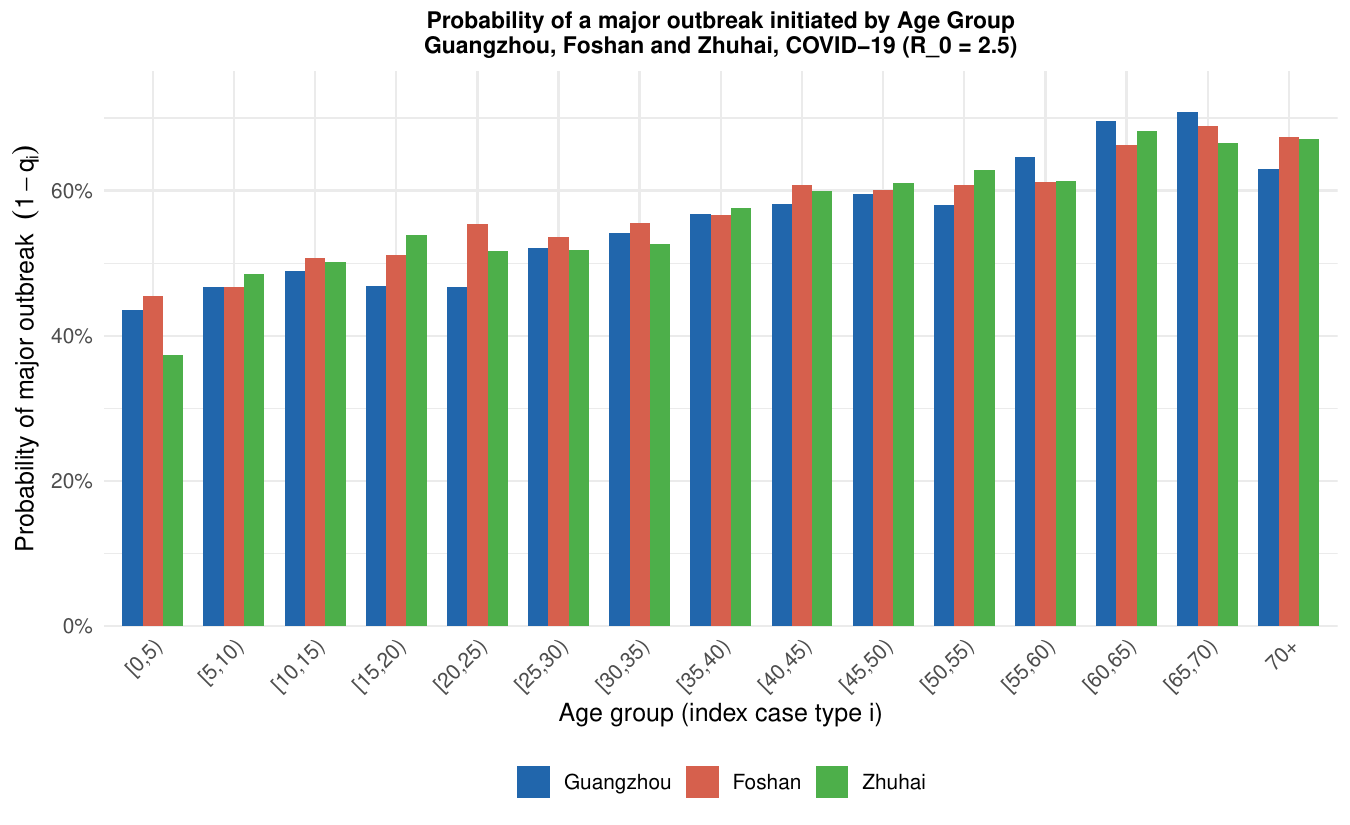}
    \caption{Probability of a major outbreak by seeding age group (initiations) across Guangzhou, Foshan and Zhuhai (COVID-19, $\mathcal{R}_0=2.5$). Bars show the predicted probability of a major outbreak if the epidemic is initiated by one infectious individual in that age group. The approximate branching process formulas in \eqref{pgfs} and \eqref{simpliedpgfs} are applied by combining
    the empirical contact matrices in three cities and age-stratified epidemiological parameters. Colors denote city: Guangzhou (blue), Foshan (orange), Zhuhai (green).}
    \label{fig:comparison_probs_outbreak}
\end{figure}

The probabilities of a major outbreak initiated from one infectious individual from all age groups in three cities are 
illustrated in Figure \eqref{fig:comparison_probs_outbreak}. 

\begin{figure}[htbp]
    \centering
    \includegraphics[width=0.8\linewidth]{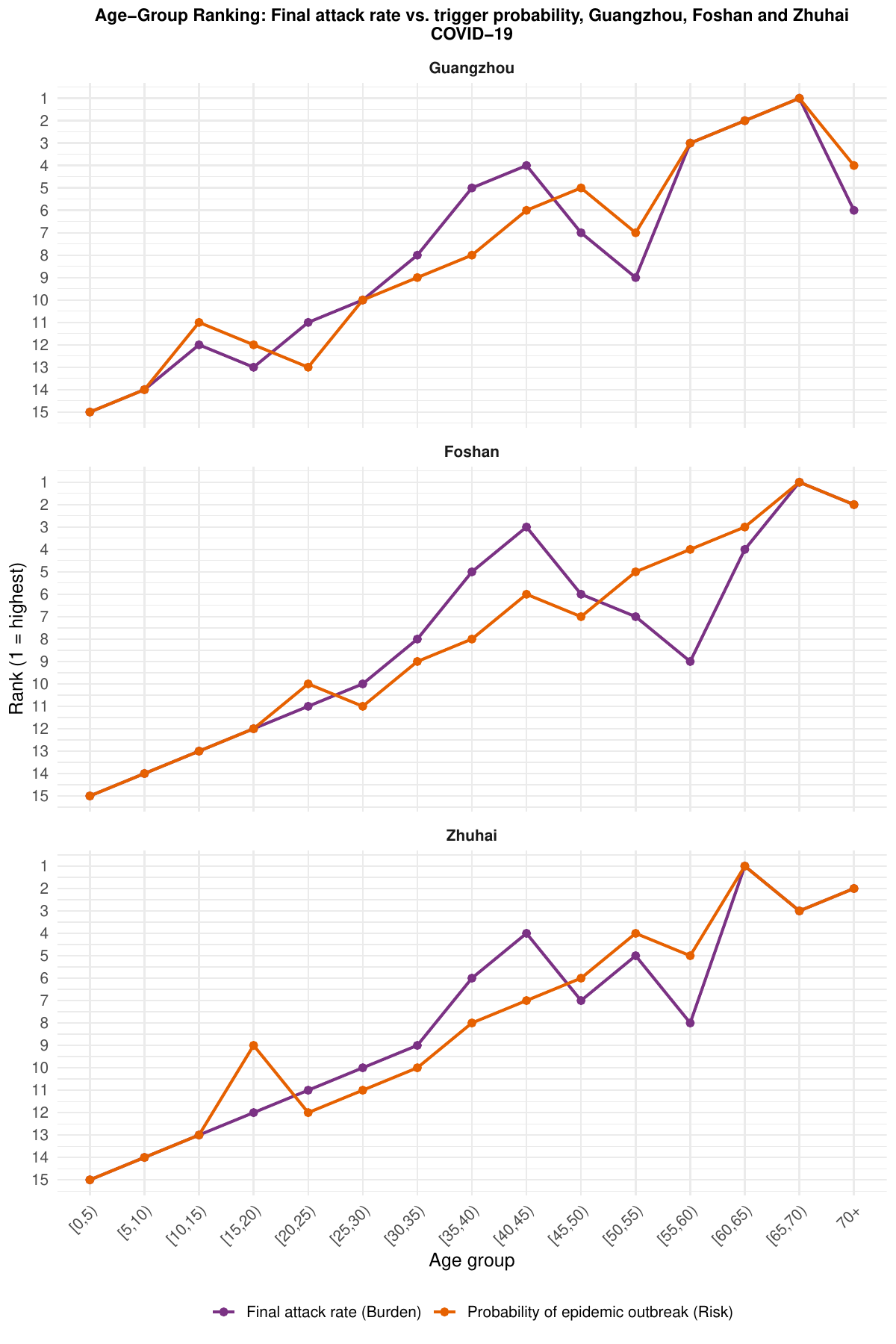}
    \caption{Ranking of age groups across three cities, based on the final attack rates and the probabilities of a major epidemic outbreak.}
    \label{fig:rank_probs_attackrates}
\end{figure}

Bars in Figure \eqref{fig:comparison_probs_outbreak} show the analytical outbreak trigger probability
$1 - q_k$ for a single index case of age group $k$,
computed from the multi-type branching-process
approximation of the CTMC model (equations~\eqref{stochasticSIR}
and~\ref{simpliedpgfs}) using the empirical contact matrices $p_{ij}$ for each city. The probability of a major outbreak increases monotonically overall with
age across all three cities, reflecting that the age-increasing infectious period $1/\gamma_i$ in the parameterization for COVID-19 is the dominant factor in predicting the early-stage epidemic outbreak. The longer assumed infectious periods among older age groups yield the increased expected number of secondary infections generated by one initial case. Therefore, the predicted probability of major outbreak is also increased. 
For COVID-19, the probability ranking is generally consistent with the predicted attack-rate ranking. The comparisons of ranks of final attack rates and probabilities of epidemic outbreak across three cities are presented in Figure \eqref{fig:rank_probs_attackrates}.

\medskip

Across all three cities, both metrics demonstrate a pronounced, monotonically increasing trend from pediatric groups to older adults. This alignment confirms that for adult-driven, high-severity respiratory pathogens, burden-targeted and trigger-targeted surveillance strategies coincide for epidemics such as COVID-19. However, a notable divergence occurs within working-age adult cohorts: across Guangzhou, Foshan, and Zhuhai, the $[40,45)$ age group exhibits a disproportionately higher rank in cumulative attack rate, compared to its early-stage outbreak trigger probability. This indicates that while middle-aged adults accumulate substantial infection burden through high workplace and community exposure, individual introductions within this cohort are less efficient at sustaining early-stage stochastic branching chains than introductions among older, highly susceptible individuals. More importantly, the ranking profiles for COVID-19 show high structural consistency across all three cities. The simulation demonstrates that macro-level adult transmission networks are robust to municipal-level contact variations.

\medspace

\subsection{The long-term and short-term analysis of Hand-Foot-Mouth disease (HFMD) transmission}

The Hand-Foot-Mouth disease (HFMD), caused by several enteroviruses including enterovirus A71 and coxsackievirus A16, is a childhood infection of major public health importance in southern China. A systematic study on age patterns and transmission characteristics showed that, the force of infection rises rapidly below $3$ to $4$ years of age and peaks at ages $3$ to $5$, with a smaller secondary peak at ages $25$ to $39$ that has been attributed to parents infected by children (\cite{Zhao2016hfmd}). The immunity against HFMD accumulates through childhood, so the susceptibility is concentrated in the youngest ages. It was investigated in \cite{Li2013hfmd}, that seroprevalence of coxsackievirus A16 steadily increased from $46.7\%$ at $1$ year group to $88\%$ among $5$–$20$ year old age group in Guangdong province before the $2010$ epidemic. This makes HFMD an ideal case study for demonstrating the importance of age-specific heterogeneity in susceptibility. Moreover, the contact patterns in all three cities show that young children mostly contact with other young children with same age. It is interesting to investigate that whether the age groups of young children that bear the burden are also the group most able to trigger a major outbreak. This is particularly relevant to the design of interventions. 

We model HFMD transmission as a single generic pathogen transmitted through person-to-person contact. Serotype-specific dynamics, different pathways of symptomatic infections and asymptomatic infections and indirect routes are not considered. Prior immunity for older groups is included in relative susceptibility $u_i$ (\cite{Li2013hfmd}). The baseline $u_i$ is set for under-$5$ age group and $u_i$ for all other groups are estimated. $\mathcal{R}_0$ is set to be $1.5$, which is in line with the effective reproduction numbers of $1.4$-$1.6$ estimated for the spring peak in China reported in \cite{Wang2011}. The infectious periods and relative infectivities for all age groups are assumed to be uniform across ages due to lack of direct evidence. Incubation periods are assumed based on data in \cite{Yang2017hfmd}. The detailed age-stratified parameters are listed in Table~\ref{tab:hfmd_params} in Appendix~\ref{app:figuresandtables}. It is emphasized that, the final attack rates for all age groups refer to infections, not symptomatic infections. It is evident that substantial proportions of infections across all age groups are asymptomatic ones, and the proportions are age related.

\begin{figure}[htbp]
    \centering
    \includegraphics[width=\linewidth]{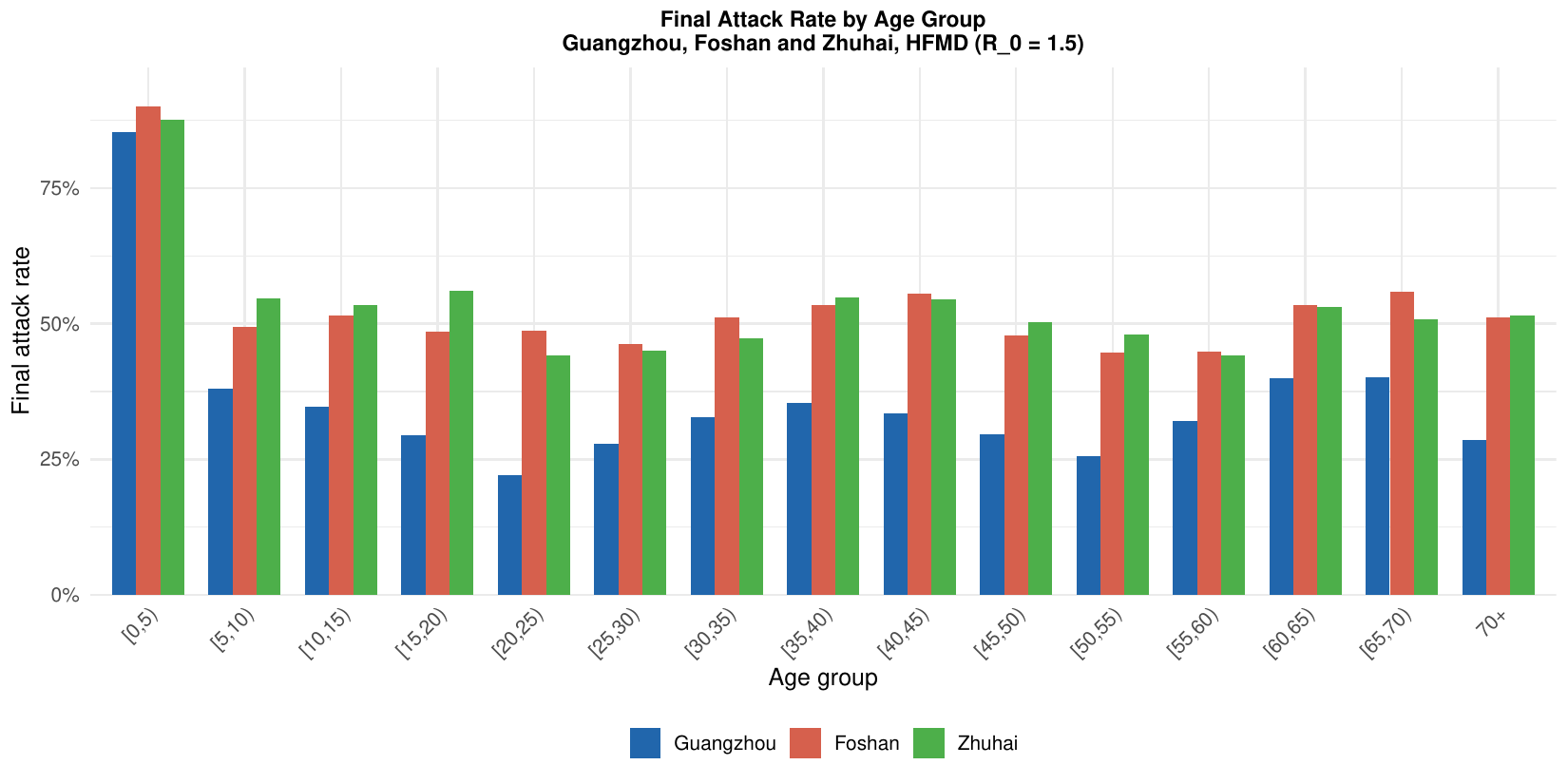}
    \caption{Final attack rates across Guangzhou, Foshan and Zhuhai (HFMD, $\mathcal{R}_0=1.5$). Bars show the estimated attack rate in each age group. The final size relations in \eqref{final_size_relation} are applied by combining the empirical contact matrices in three cities and age-stratified epidemiological parameters. Colors denote city: Guangzhou (blue), Foshan (orange), Zhuhai (green).}
    \label{fig:comparison_attack_rates_HFMD}
\end{figure}

Figure~\ref{fig:comparison_attack_rates_HFMD} illustrates the estimated final attack rates for all age groups in three cities. The under-$5$ age group is the most affected in all three cities, with attack rates of $85.3\%$ in Guangzhou, $90.0\%$ in Foshan, and $87.5\%$ in Zhuhai. For other age groups, the final attack rates are with range $22.1\%$-$40.2\%$, $44.7\%$-$55.8\%$ and $44.1\%$-$56.0\%$, respectively. A closer comparison of the final attack rates across the three cities reveals distinct structural variations shaped by underlying demographic and different contact patterns. While Foshan and Zhuhai exhibit relatively broad distributions of infection across older cohorts (with adult attack rates ranging between $44.1\%$ and $56.0\%$), Guangzhou displays a sharper structural decoupling between the heavily impacted under-$5$ cohort ($85.3\%$) and all other age groups. This amplification in Guangzhou is driven by higher age-assortative mixing for under-$5$ age group. The comparison of predictions by using age-structured model and corresponding homogeneous mixing model is presented in Figure~\ref{fig:comparison_attack_rates_homogenous_heterogeneous_outbreak_outbreak_HFMD} in Appendix~\ref{app:figuresandtables}.

\begin{figure}[htbp]
    \centering
    \includegraphics[width=\linewidth]{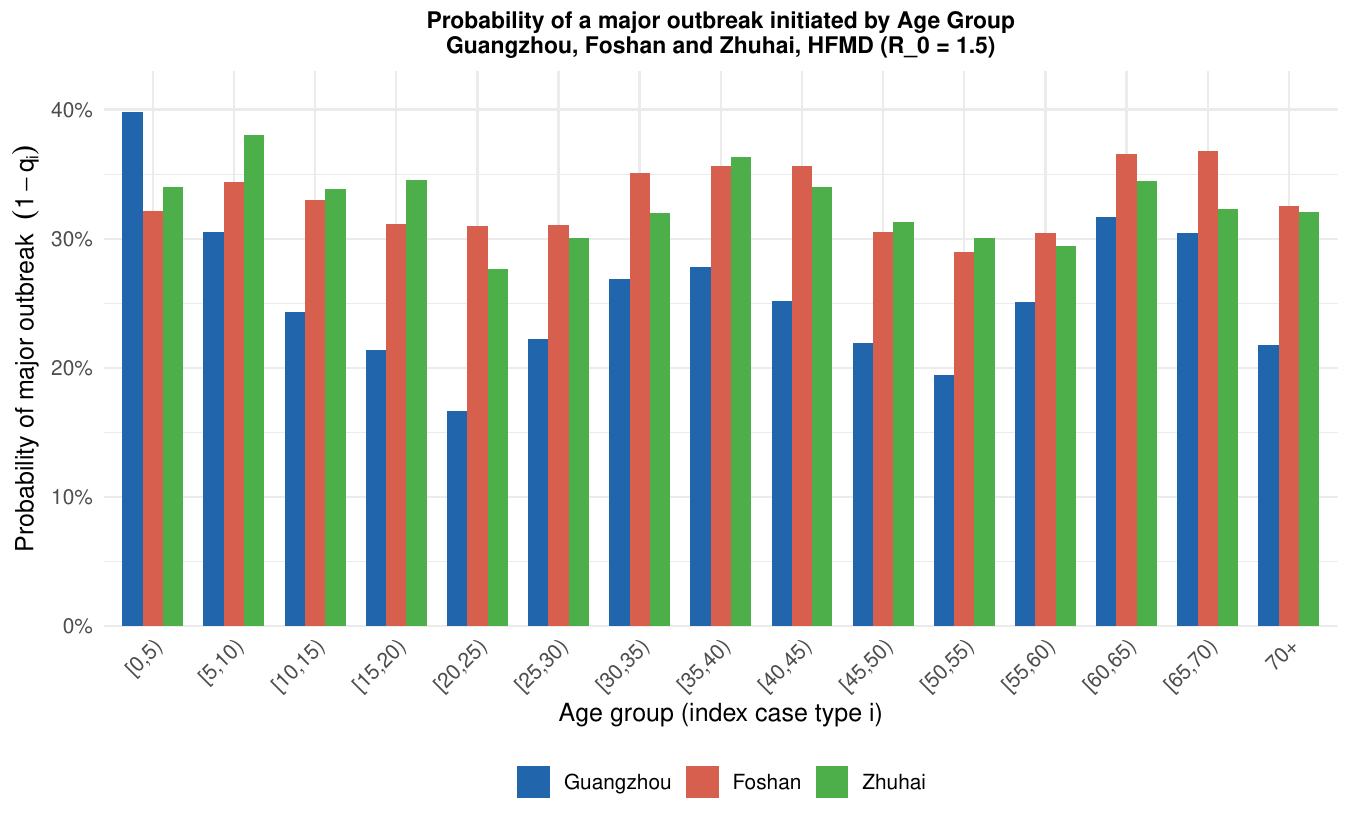}
    \caption{Probability of a major outbreak by seeding age group (initiations) across Guangzhou, Foshan and Zhuhai (HFMD, $\mathcal{R}_0=1.5$). Bars show the predicted probability of a major outbreak if the epidemic is initiated by one infectious individual in that age group. The Approximate branching process formulas in \eqref{pgfs} and \eqref{simpliedpgfs} are applied by combining the empirical contact matrices in three cities and age-stratified epidemiological parameters. Colors denote city: Guangzhou (blue), Foshan (orange), Zhuhai (green).}
    \label{fig:comparison_probs_outbreak_outbreak_HFMD}
\end{figure}

Early-stage stochastic outbreak risks for age groups in the three cities are evaluated by branching process analysis ($1-q_i$). 
Figure~\ref{fig:comparison_probs_outbreak_outbreak_HFMD} presents the predicted probability of a major outbreak initiated by a single infectious individual in each age group. The age-specific patterns differ across cities. In Guangzhou, the under-five group has the highest outbreak-trigger probability, consistent with its exceptionally high final attack rate. In contrast, in Foshan and Zhuhai, the under-five group does not rank first in outbreak-trigger probability, despite having the highest final attack rate in both cities. This contrast illustrates that the age group bearing the greatest long-term infection burden is not necessarily the group whose introduction is most likely to initiate a major outbreak. Outside the under-5 cohort, Guangzhou exhibits significantly lower trigger probabilities across all other age groups relative to Foshan and Zhuhai. In addition, older age groups also exhibit elevated trigger probabilities, highlighting the potential importance of inter-generational contact patterns in early-stage transmission.

\begin{figure}[htbp]
    \centering
    \includegraphics[width=0.8\linewidth]{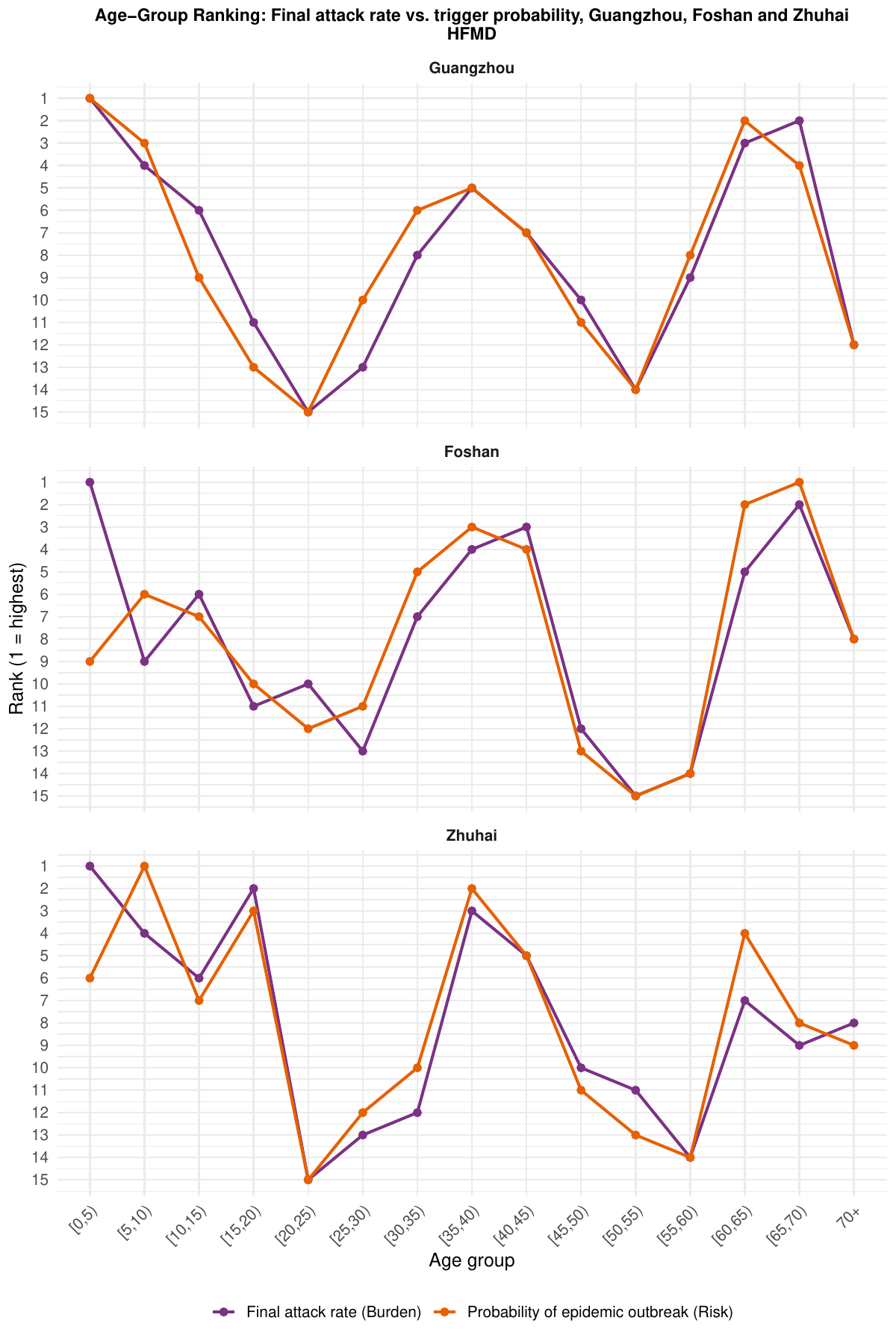}
    \caption{Ranking of age groups across three cities, based on the final attack rates and the probabilities of a major Hand-Foot-Mouth disease (HFMD) outbreak.}
    \label{fig:rank_probs_attackrates_HFMD}
\end{figure}

Figure~\ref{fig:rank_probs_attackrates_HFMD} ranks the $15$ age groups by final attack rate (burden) and by trigger probability (risk). The two rankings converge over most of the age range, with Spearman correlations of $0.94$, $0.82$ and $0.90$ in Guangzhou, Foshan and Zhuhai. The same shape appears in all three cities: both ranks are high at ages $35$-$45$ and $60$-$70$ in Guangzhou and Foshan, and lowest at ages $50$-$60$ and $20$-$25$ in Guangzhou and Zhuhai. 
The two rankings are broadly similar across most age groups, suggesting that the age-specific contact structure plays an important role in shaping both cumulative infection burden and early-stage outbreak risk. In particular, age groups with substantial connections to other groups may rank highly on both measures, consistent with the parent–child and grandparent–child contact patterns observed in the empirical matrices. However, the rankings are not identical, because cumulative infection burden and outbreak-trigger probability reflect different aspects of transmission dynamics, and the model also incorporates age-specific susceptibility. The most notable divergence occurs in the under-five age group. This group ranks first in final attack rate in all three cities, reflecting its higher assumed susceptibility, but ranks first, ninth, and sixth in outbreak-trigger probability in Guangzhou, Foshan, and Zhuhai, respectively. The differences across cities suggest that the ability of an introduction in this group to initiate a major outbreak depends not only on its susceptibility but also on how its contacts connect it to susceptible individuals within and outside the age group. School-age children $[5, 10)$ rank first in outbreak-trigger probability in Zhuhai, compared with fourth in final attack rate, while the $[60, 65)$ age group ranks higher for outbreak-trigger probability than for final attack rate in both Foshan and Zhuhai (second versus fifth, and fourth versus seventh, respectively). 
Excluding the under-five group, the Spearman rank correlations are $0.93$, $0.94$, and $0.94$, indicating a strong association between the two rankings among the remaining age groups.

\medskip

Several limitations should be considered when interpreting these results. The model does not explicitly distinguish symptomatic infections and asymptomatic infections. The age-specific asymptomatic proportions are also not incorporated. Such heterogeneity can affect the interpretation of age-specific infection and transmission patterns of HFMD, for which asymptomatic or mild infections are relevant. 

\medskip

\subsection{Implications for public-health prioritization}
The comparison between long-term epidemic burden and early-stage outbreak risk provides complementary information for public-health planning. The age-specific final attack rate, $\pi_i$, characterizes the cumulative infection burden predicted, whereas $1-q_i$ measures the probability that an introduction into age group $i$ initiates a major outbreak. The two quantities therefore represent different epidemiological objectives. A burden-oriented perspective may help identify age group(s) likely to experience a large number of infections, while the trigger probability may help identify groups for which early detection and investigation of introductions warrant particular attention.

The COVID-19 and HFMD examples illustrate the value of considering both measures. For COVID-19, the age patterns of final attack rates and major-outbreak probabilities are broadly aligned. For HFMD, the high attack rate in the under-five group contrasts with the variation in its relative outbreak-trigger risk across the three cities. These results suggest that public-health priorities informed by cumulative infection burden may differ from those informed by the probability of initiating a major outbreak. The extent of the difference depends on the pathogen-specific epidemiological assumptions, age-dependent susceptibility and infectiousness, and the contact patterns of the population under study.

These findings support the use of complementary epidemiological indicators when planning age-specific surveillance and preparedness. However, the two measures alone do not determine an optimal allocation of vaccination, testing, or healthcare resources. Evaluating the effectiveness of specific interventions would require extending the model to represent the intervention mechanisms and defining the corresponding public-health objectives. Such intervention-specific analyses are important directions for future work. 

\medspace
\section{Conclusion and future work}
In this study, we combine deterministic age-structured epidemic models with a CTMC-based branching process analysis to investigate the relationship between age-specific cumulative infection burden and early-stage outbreak-trigger probability. Using contact matrices generated from empirical social contact data in Guangzhou, Foshan and Zhuhai, we examined two disease scenarios, COVID-19 and hand, foot, and mouth disease (HFMD), to assess whether the age groups experiencing the greatest cumulative infection burden are also those from which an initial infection is most likely to trigger a major outbreak in the population.

The simulated results demonstrate that the relationship between these two measures depends on the epidemiological parameters and overall population contact pattern. For the case study of COVID-19, the predicted outbreak-trigger probability increases monotonically with age across all three cities, primarily reflecting the assumed age-dependent lengths of infectious periods. Differences in the magnitude of this probability across cities may also be associated with their distinct age-specific contact patterns. In contrast, the predicted final attack rate generally increases with age but shows a slight decline in the oldest age group. For the case study of HFMD, the differences between the two measures are more pronounced, particularly for the under-five age group. Its higher assumed susceptibility leads to the highest final attack rate in all three cities. However, its ranking in outbreak-trigger probability varies substantially across three cities, reflecting differences in their age-specific contact patterns. In Guangzhou, the relatively high degree of within-group assortative mixing among young children, together with their higher susceptibility, contributes to the particularly high outbreak-trigger probability of this group. In contrast, the lower proportions of within-group contacts in Foshan and Zhuhai are associated with relatively lower trigger-probability rankings for the under-five group. Despite these differences, the rank correlations between the two measures remain high overall, indicating a strong association between long-term cumulative infection burden and early-stage outbreak risk across age groups. Together, these findings illustrate that the age groups bearing the greatest cumulative infection burden do not necessarily have the highest probability of triggering a major epidemic outbreak. The two measures therefore address different epidemiological objectives and may provide complementary information for surveillance and intervention planning in public health management.

Several limitations should be considered when interpreting the results. First, the epidemiological parameters and assumptions differ between the two disease scenarios and may not capture the full range of biological heterogeneity within each disease. In particular, the mathematical models for HFMD transmission do not explicitly distinguish symptomatic and asymptomatic infections or incorporate age-specific asymptomatic proportions. Second, the empirical social contact data used in this study were collected in 2016, approximately a decade ago. Population contact patterns may have changed substantially since then, particularly following the COVID-19 pandemic. Studies comparing social contact patterns before and after the pandemic, such as the study conducted in Switzerland (\cite{Jarvis2024,Reichmuth2024}), have highlighted the potential for changes in population mixing behavior. Therefore, the contact matrices used in the present study may not fully reflect current contact patterns in Guangzhou, Foshan, and Zhuhai. If feasible, new social contact surveys in these cities would be valuable for assessing changes over time and evaluating how such changes may affect the predicted age-specific infection burden and outbreak-trigger probabilities.

Future work could extend the framework in several directions. Incorporating more detailed age-specific epidemiological characteristics, including heterogeneity in clinical presentation and transmission, could improve the representation of disease-specific dynamics. Another direction is to develop a statistical framework for estimating the parameters of preferred mixing patterns using individual-level social contact data. Such an approach could be used to quantify and compare the degree of age-assortative mixing across different geographical locations, providing a more systematic understanding of how population contact structures vary across regions. Furthermore, it would be valuable to investigate the mechanisms underlying the relationship between age-specific cumulative infection burden and early-stage outbreak-trigger probability. This could be achieved through controlled numerical experiments in which contact patterns are varied while other epidemiological parameters are held fixed. Such analyses would help assess how changes in contact structure influence the two epidemiological outcomes and provide insights that could inform the design of public health interventions.

\medspace
\ \\
\noindent {\bf Declaration of Competing Interest:}
None.

\medspace

%
%
%

\section*{CRediT authorship contribution statement}

Sicong Zhang: Conceptualization, Funding acquisition, Investigation,
Methodology, Project administration, Resources.

Mingyuan Wu: Conceptualization, Data curation, Investigation,
Methodology, Writing -- review \& editing.

Jian Yang: Conceptualization, Data curation, Investigation,
Methodology, Writing -- review \& editing.

Jianpeng Xiao: Conceptualization, Data curation, Funding acquisition.

Fan Bai: Conceptualization, Data curation, Formal analysis,
Funding acquisition, Investigation, Methodology, Project administration,
Resources, Software, Supervision, Validation, Visualization,
Writing -- original draft, Writing -- review \& editing.

\medspace

\noindent {\bf Acknowledgments:}
The work was supported by the National Natural Science Foundation of China ($82373636$).

\medspace

\noindent {\bf Declaration of generative AI and AI-assisted technologies in the manuscript preparation process:}
During the preparation of this work, the authors used ChatGPT (OpenAI) and Claude (Anthropic) to assist with improving the language, clarity, and readability of the manuscript. The authors reviewed and edited the AI-assisted content as necessary and take full responsibility for the accuracy and integrity of the published work.

\clearpage
\newpage

\bibliography{bibfile}
\bibliographystyle{plain}
\clearpage

\appendix
 
\section{The final size relation of age-structured epidemic model}
\label{app:derivationsDet}
Dividing both sides of the first equation in \eqref{mainmodel} by $S_i$,
\begin{equation*}
\dfrac{S_{i}^{\prime}}{S_i} = -\beta_i\sum\limits_{j=1}^{n}\epsilon_{ij}p_{ij}\displaystyle\frac{I_j}{N_j},    
\end{equation*}
integrating from $t=0$ to $t=\infty$,
\begin{equation}\label{relationApp1}
    \log S_i(\infty) - \log S_i(0) = -\beta_i\sum\limits_{j=1}^{n}\epsilon_{ij}p_{ij}\displaystyle\frac{1}{N_j}\int_{0}^\infty I_j(t)dt.
\end{equation}
Integrating the fourth equation in \eqref{mainmodel} from $t=0$ to $t=\infty$,
\begin{equation*}
    R_i(\infty)-R_i(0) = \gamma_i \int_{0}^\infty I_i(t)dt.
\end{equation*}
It is reasonable to assume that $R_i(0)=0$, and as $t \rightarrow \infty$, both $E_j(\infty)=I_j(\infty)=0$, so $R_i(\infty) = N_i -S_i(\infty)$. Substituting it into previous equation,
\begin{equation}\label{relationApp2}
  \int_{0}^\infty I_i(t)dt = \dfrac{N_i -S_i(\infty)}{\gamma_i}.  
\end{equation}
Substituting equation~\ref{relationApp2} into equation~\ref{relationApp1} to obtain
\begin{equation*}
\log S_i(\infty) - \log S_i(0) =  -\beta_i\sum\limits_{j=1}^{n}\epsilon_{ij}p_{ij}\displaystyle\frac{N_j-S_j(\infty)}{\gamma_j N_j}.  
\end{equation*}
Considering that $N_i \approx S_i(0)$, we rearrange the above equation to have
\begin{equation}\label{finalrelation}
 \log \dfrac{N_i}{S_i(\infty)} = \beta_i\sum\limits_{j=1}^{n}\epsilon_{ij}p_{ij}\displaystyle\frac{N_j-S_j(\infty)}{\gamma_j N_j}.     
\end{equation}
The final size relation \eqref{finalrelation} is able to estimate the ultimate burden in each age group. It is noticed that the incubation rate $\delta_i$ does not appear in the final size relation, implying that this parameter does not affect the final epidemic outcome, but only the timing of epidemic waves.

\newpage
\section{Generating the probability of a minor/major outbreak}
\label{app:derivationSto}
\begin{theorem}
	In the multi-type branching process approximation, assume that there is an exposed individual or an infectious individual in the sub-group of $i \in \{1, 2,\cdots ,n\}$. If the basic reproduction number is $\mathcal{R}_0 \leq 1$, then the probability of a minor outbreak is $\mathbb P_{\textit{minor}}=1$, and if $\mathcal{R}_0 > 1$, the probability of a minor outbreak is $\mathbb P_{\textit{minor}} = q_i$. $q_i$ is defined as the minimal non-negative fixed point of equation \eqref{pgfs}.
\end{theorem}

\begin{proof}
We define $p_{(i_1,i_2,\cdots,i_{2n-1},i_{2n}),(j_1,j_2,\cdots,j_{2n-1},j_{2n})}$ as the transition probability for $$E_1(t),I_1(t),\cdots,E_n(t),I_n(t)$$ in the branching process approximation. We have
\begin{equation*}
p_{(i_1,\cdots,i_{2n}),(j_1,\cdots,j_{2n})}(t) = \mathbb{P}((E_1(t),I_1(t),\cdots,I_n(t))=(j_1,\cdots,j_{2n}) \mid (E_1(0),I_1(0),\cdots,I_n(0))=(i_1,\cdots,i_{2n})).    
\end{equation*} 
The pgf (probability generating function) can be written as,
\begin{equation}\label{pgdnotion1}
G_{(i_1,\cdots,i_{2n})}(u_1,u_2,\cdots,u_{2n},t) =  \displaystyle\sum\limits_{j_1,j_2,\cdots,j_{2n}} p_{(i_1,\cdots,i_{2n}),(j_1,\cdots,j_{2n})}(t)u_1^{j_1}u_2^{j_2}\cdots u_{2n}^{j_{2n}}.   
\end{equation}
The independence assumption yields that the pgf $G_{(i_1,i_2,\cdots,i_{2n})}$ can be defined as 
\begin{equation}\label{pgdnotion2}
\begin{aligned}
G_{(i_1,i_2,\cdots,i_{2n})}(u_1,u_2,\cdots,u_{2n},t) =& [G_{(1,0,\cdots,0)}(u_1,u_2,\cdots,u_{2n},t)]^{i_1} \\
& \times [G_{(0,1,\cdots,0)}(u_1,u_2,\cdots,u_{2n},t)]^{i_2} \\
& \times \cdots\cdots \\
& \times [G_{(0,0,\cdots,1)}(u_1,u_2,\cdots,u_{2n},t)]^{i_{2n}}.
\end{aligned}
\end{equation}
We next formulate the backward Kolmogorov differential equations for the branching process approximation of $E_1(t), I_1(t), \cdots, E_n(t), I_n(t)$, and for the sake of simplicity, we denote 
\begin{equation*}
p_{a,b}(t) := p_{(i_1, i_2, \cdots,i_{2n}),(j_1,j_2, \cdots,j_{2n})}(t).    
\end{equation*}
The derived backward Kolmogorov differential equations are as follows,
\begin{equation}\label{backwardKDE}
\begin{aligned}
\dfrac{dp_{a,b}(t)}{dt} &= \beta_1 \displaystyle\sum\limits_{j=1}^{n}\epsilon_{1j}p_{1j}\frac{i_{2j} N_1}{N_j}p_{(i_1+1,i_2,\cdots,i_{2n}),b} + \beta_2 \displaystyle\sum\limits_{j=1}^{n}\epsilon_{2j}p_{2j}\frac{i_{2j} N_2}{N_j}p_{(i_1,i_2,i_3+1\cdots,i_{2n}),b} \\
&+ \cdots\cdots + \beta_n \displaystyle\sum\limits_{j=1}^{n}\epsilon_{nj}p_{nj}\frac{i_{2j} N_n}{N_j}p_{(i_1,\cdots,i_{2n-1}+1,i_{2n}),b} \\
&+ \delta_1 i_1 p_{(i_1-1,i_2+1,\cdots,i_{2n}),b} + \cdots\cdots+ \delta_n i_{2n-1} p_{(i_1,\cdots,i_{2n-1}-1,i_{2n}+1),b}\\
&+\gamma_1 i_2 p_{(i_1,i_2-1,\cdots,i_{2n}),b}
+ \cdots\cdots + \gamma_n i_{2n} p_{(i_1,\cdots,i_{2n}-1),b} \\
&-(\displaystyle\sum\limits_{i=1}^{n}\sum\limits_{j=1}^{n}\beta_i \epsilon_{ij} p_{ij} i_{2j}\frac{N_i}{N_j}+\displaystyle\sum\limits_{i=1}^{n}\delta_i i_{2i-1}+\displaystyle\sum\limits_{i=1}^{n}\gamma_i i_{2i})p_{a,b}.
\end{aligned}
\end{equation}
If we consider the case where $i_1 > 0$ and $i_j = 0$ for all $j > 1$, then we have
\begin{equation}\label{diffeqn}
\begin{aligned}
\dfrac{\partial G_{(1,0,\cdots,0)}}{\partial t} &= \frac{1}{i_{1}[G_{(1,\cdots,0)}]^{i_{1}-1}}\frac{\partial G_{(i_1,\cdots,0)}}{\partial t} \\
&=\frac{1}{i_{1}[G_{(1,\cdots,0)}]^{i_{1}-1}}\displaystyle\sum_{b}\frac{dp_{(i_{1},\cdots,0),b}}{dt}u_{1}^{j_{1}}\cdots u_{2n}^{j_{2n}} \\
&=\frac{1}{i_{1}[G_{(1,\cdots,0)}]^{i_{1}-1}}\displaystyle\sum_{b}[\delta_1 i_1 p_{(i_1-1,1,\cdots,0),b}  - \delta_1 i_1 p_{(i_1,0,\cdots,0),b}] \\
&=\delta_1 G_{(0,1,\cdots,0)} - \delta_1 G_{(1,0,\cdots,0)}.
\end{aligned}    
\end{equation}
The steady state of \eqref{diffeqn} can be expressed as the fixed points of the first equation in \eqref{pgfs}.
\begin{equation*}
G_{(1,0,\cdots,0)} = G_{(0,1,\cdots,0)}.
\end{equation*}
Because $G_{(1,0,\cdots,0)}(0,\cdots,0,t) = p_{(1,0,\cdots,0),(0,0,\cdots,0)}(t)$ and $G_{(0,1,\cdots,0)}(0,\cdots,0,t) = p_{(0,1,\cdots,0),(0,0,\cdots,0)}(t)$, two probabilities of a minor outbreak by introducing one exposed individual of age group $1$ or one infectious individual of age group $1$ are equal.

We next consider the case where $i_2 >0$ and all others are set to be $0$,
\begin{equation}\label{diffeqn1}
\begin{aligned}
\dfrac{\partial G_{(0,1,\cdots,0)}}{\partial t} &= \frac{1}{i_{2}[G_{(0,1,\cdots,0)}]^{i_{2}-1}}\frac{\partial G_{(0,i_2,\cdots,0)}}{\partial t} \\
&=\frac{1}{i_{2}[G_{(0,1,\cdots,0)}]^{i_{2}-1}}\displaystyle\sum_{b}\frac{dp_{(0,i_{2},\cdots,0),b}}{dt}u_{1}^{j_{1}}\cdots u_{2n}^{j_{2n}} \\
&=\frac{1}{i_{2}[G_{(0,1,\cdots,0)}]^{i_{2}-1}}\displaystyle\sum_{b}[\beta_1 \epsilon_{11}p_{11}i_2 p_{(1,i_2,0,\cdots,0),b} + \cdots\cdots \\
&+\beta_n \epsilon_{n1}p_{n1}i_2\dfrac{N_n} {N_1}p_{(0,i_2,\cdots,1,0),b}] + \gamma_1 i_2 p_{(0,i_2-1,\cdots,0),b} \\
&- (\sum_{i=1}^n \beta_i \epsilon_{i1}p_{i1}i_2 \dfrac{N_i}{N_1} + \gamma_1 i_2)p_{(0,i_2,0,\cdots,0),b} \\
&= \beta_1\epsilon_{11}p_{11}G_{(1,0,\cdots,0)}G_{(0,1,\cdots,0)}+\cdots+\beta_n\epsilon_{n1}p_{n1}\dfrac{N_n}{N_1}G_{(0,1,\cdots,0)}G_{(0,0,\cdots,1,0)} \\
&+\gamma_1 - (\sum_{i=1}^n \beta_i \epsilon_{i1}p_{i1} \dfrac{N_i}{N_1} + \gamma_1)G_{(0,1,\cdots,0)}.
\end{aligned}
\end{equation}
The steady state of \eqref{diffeqn} can be expressed as the fixed points of the second equation in \eqref{pgfs},
\begin{equation*}
\begin{aligned}
G_{(0,1,\cdots,0)} = \dfrac{\beta_1\epsilon_{11}p_{11}G_{(1,0,\cdots,0)}G_{(0,1,\cdots,0)}+\cdots+\beta_n\epsilon_{n1}p_{n1}\dfrac{N_n}{N_1}G_{(0,1,\cdots,0)}G_{(0,0,\cdots,1,0)}+\gamma_1}{\sum_{i=1}^n \beta_i \epsilon_{i1}p_{i1} \dfrac{N_i}{N_1} + \gamma_1}   
\end{aligned}    
\end{equation*}
The proof for the rest of states is straightforward, and the analysis on stability can be found in \cite{Allen2013}.

\end{proof}

 \newpage
\section{Derivation of basic reproduction number $\mathcal{R}_0$}
\label{app:R0}
 There are $2n$ infection related compartments in the model~\ref{mainmodel},
 \begin{equation*}
     (E_1,\cdots,E_n,I_1,\cdots,I_n).
 \end{equation*}
For new infection matrix $\mathcal{F}$, we write it in block form
\begin{equation*}
    \mathcal{F} = \begin{pmatrix}
        0_{n\times n}, & \mathcal{B} \\
        0_{n\times n}, & 0_{n\times n}
    \end{pmatrix}.
\end{equation*}
Matrix $\mathcal{B}$ represents the only pathway of new infection $E\rightarrow I$, and has the element $\beta_i\epsilon_{ij}p_{ij}\dfrac{N_i}{N_j}$ on the $(i,j)$ position of $\mathcal{B}$. The transition matrix $\mathcal{V}$ has the form
\begin{equation*}
    \mathcal{V} = \begin{pmatrix}
        \operatorname{diag}(\delta_i), & 0 \\
        -\operatorname{diag}(\delta_i), & \operatorname{diag}(\gamma_i)
    \end{pmatrix}.
\end{equation*}
We thus calculate the next generation matrix 
$$\mathcal{K} = \mathcal{F}\mathcal{V}^{-1} = \begin{pmatrix}
    \mathcal{B}\operatorname{diag}(1/\gamma_i), & \mathcal{B}\operatorname{diag}(1/\gamma_i) \\
    0, & 0
\end{pmatrix}.$$ 
The basic reproduction number $\mathcal{R}_0$ is calculated as the spectral radius $\rho(\mathcal{K})$.

\newpage

\section{Supplementary figures and tables}
\label{app:figuresandtables}

\begin{table}[htbp]
\centering
\caption{Age-stratified epidemiological parameters in COVID-19 modeling}
\label{tab:covid_params}
\begin{tabular}{lcccc}
\toprule
Age group & Incubation period & Infectious period & Relative susceptibility & Relative infectivity \\
          & (days)            & (days)             & $u_i$                    & $v_j$                 \\
\midrule
{[}0,5)   & 7.5  & 6.5  & 0.50 & 1.00 \\
{[}5,10)  & 7.0  & 6.5  & 0.50 & 1.00 \\
{[}10,15) & 6.5  & 6.8  & 0.55 & 1.00 \\
{[}15,20) & 6.0  & 7.0  & 0.60 & 1.00 \\
{[}20,25) & 5.5  & 7.0  & 0.90 & 1.00 \\
{[}25,30) & 5.3  & 7.0  & 1.00 & 1.00 \\
{[}30,35) & 5.2  & 7.0  & 1.00 & 1.00 \\
{[}35,40) & 5.3  & 7.0  & 1.00 & 1.00 \\
{[}40,45) & 5.5  & 7.5  & 1.00 & 1.00 \\
{[}45,50) & 6.0  & 8.5  & 1.00 & 1.00 \\
{[}50,55) & 6.5  & 9.0  & 1.00 & 1.00 \\
{[}55,60) & 7.5  & 9.5  & 1.00 & 1.00 \\
{[}60,65) & 8.5  & 10.0 & 1.00 & 1.00 \\
{[}65,70) & 10.0 & 10.3 & 1.00 & 1.00 \\
70+       & 11.2 & 10.3 & 1.00 & 1.00 \\
\bottomrule
\end{tabular}
\end{table}

\begin{figure}[htbp]
    \centering

    \begin{subfigure}{{0.7\linewidth}}
        \includegraphics[width=\linewidth, trim=0 0 0 28pt, clip]
            {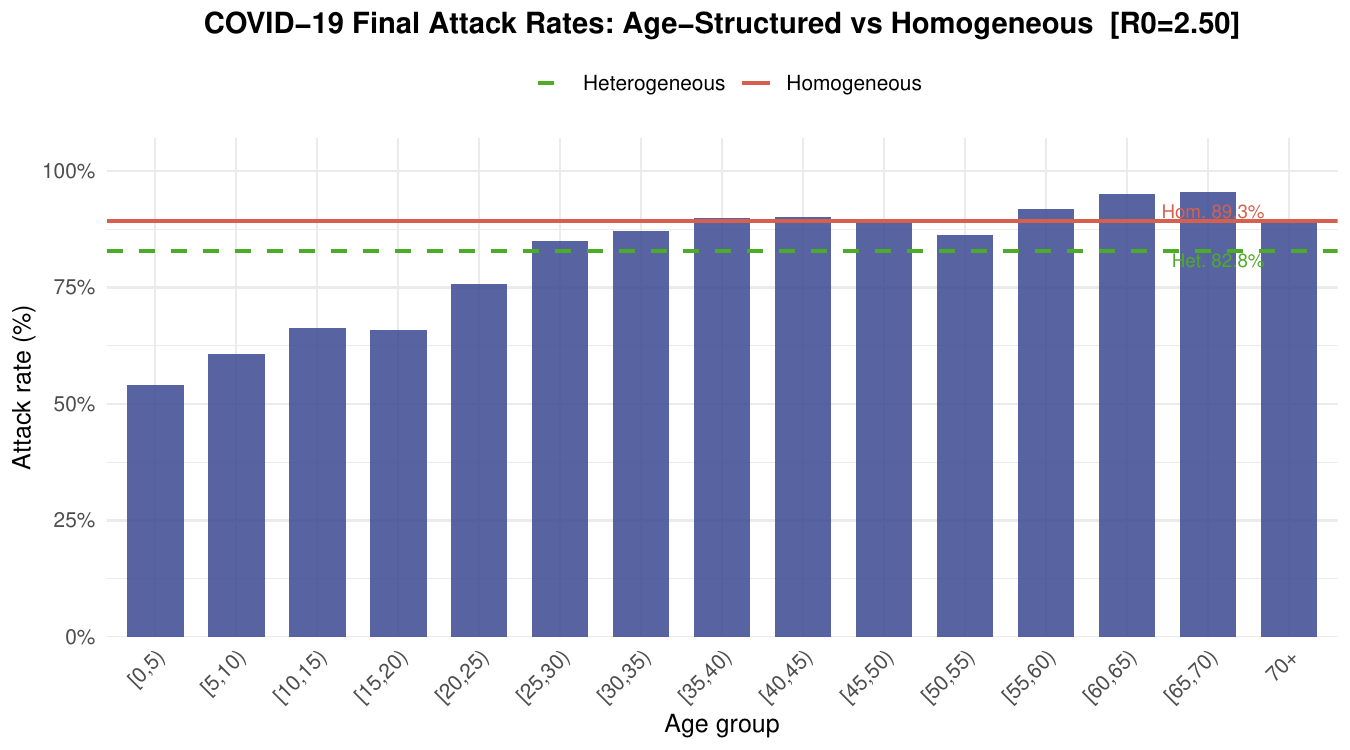}
        \caption{Guangzhou (attack ratio: 82.8\%)}
        \label{fig:ar_gz}
    \end{subfigure}

    \vspace{0.5em}

    \begin{subfigure}{0.7\linewidth}
        \includegraphics[width=\linewidth, trim=0 0 0 28pt, clip]
            {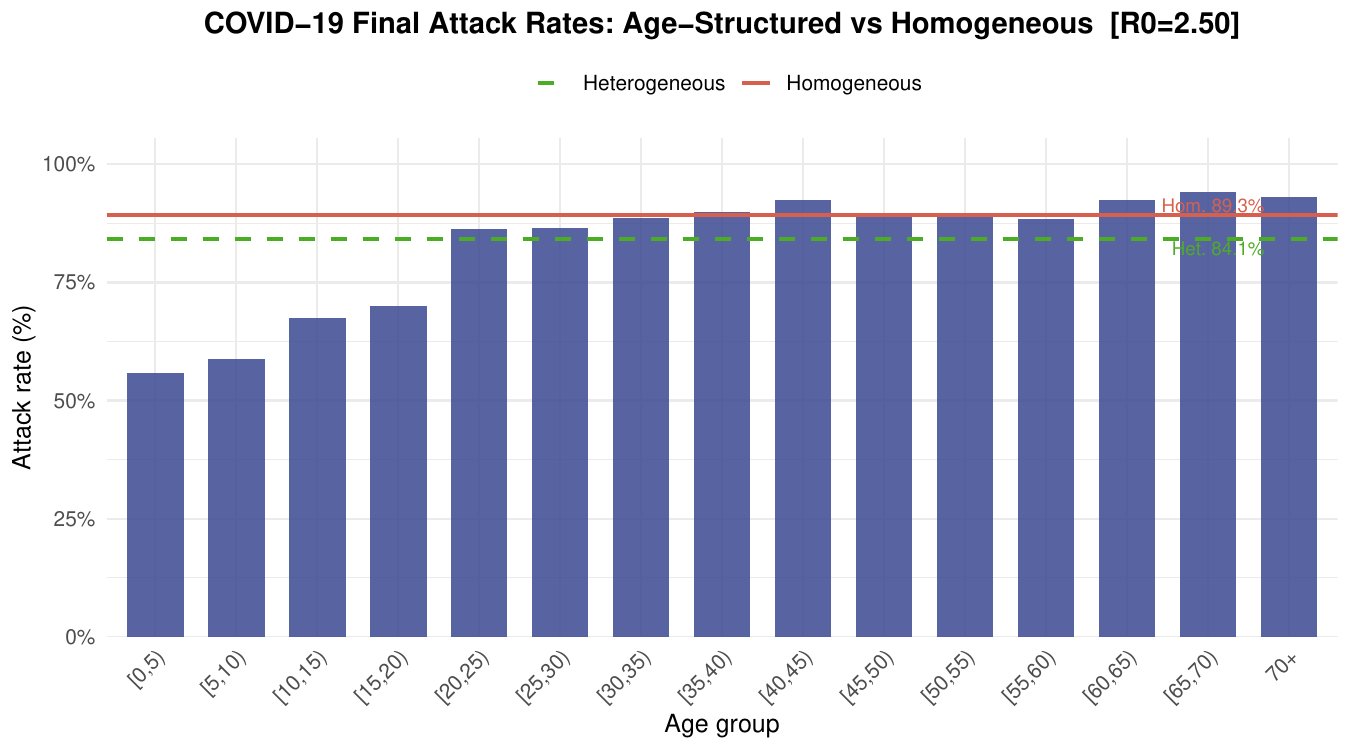}
        \caption{Foshan (attack ratio: 84.1\%)}
        \label{fig:ar_fs}
    \end{subfigure}

    \vspace{0.5em}

    \begin{subfigure}{0.7\linewidth}
        \includegraphics[width=\linewidth, trim=0 0 0 28pt, clip]
            {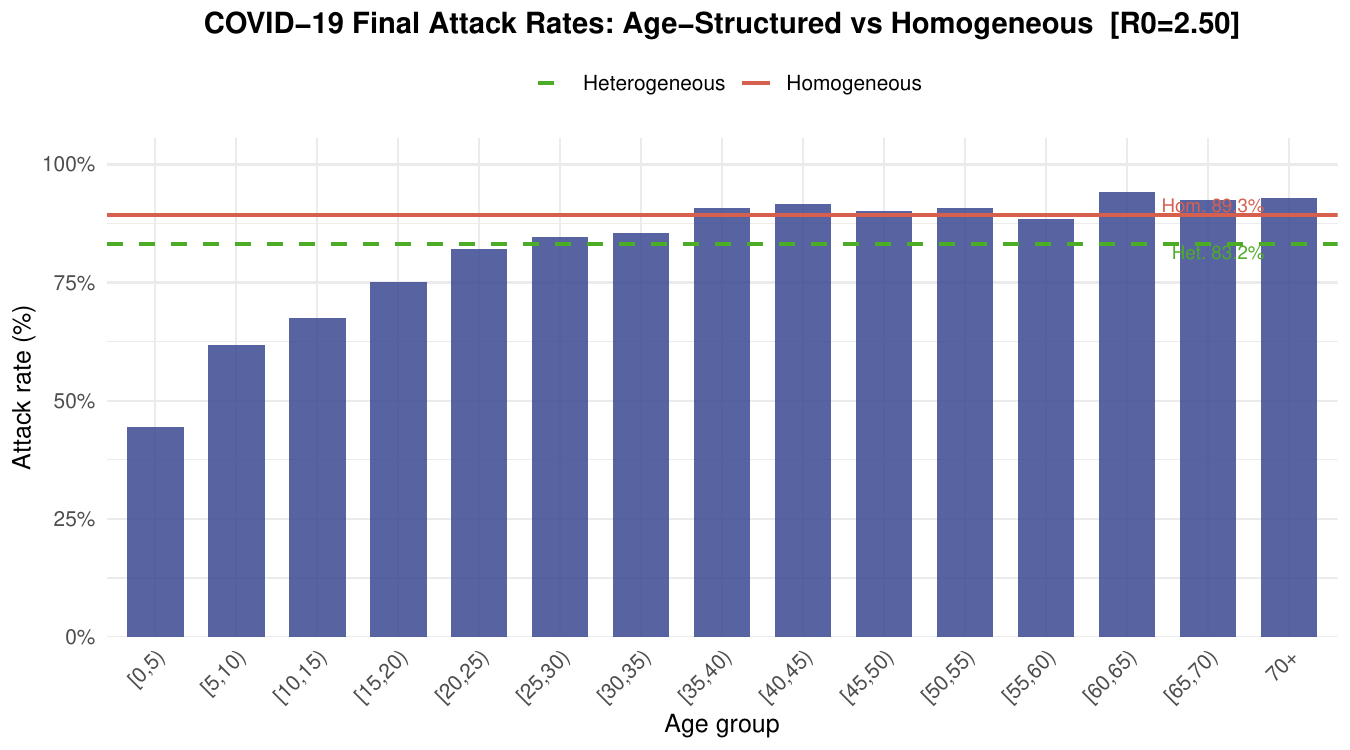}
        \caption{Zhuhai (attack ratio: 83.2\%)}
        \label{fig:ar_zh}
    \end{subfigure}

    \caption{%
        \textbf{Simulated COVID-19 final attack rates by age group across
        three cities in Guangdong Province ($R_0 = 2.50$).}
        Bars show attack rates $\pi_i$ in all age groups obtained
        from the final-size relation in
        \eqref{final_size_relation}.
        The solid red horizontal line represents the overall attack
        rate predicted by the homogeneous mixing model ($89.3\%$
        for all three cities). The dashed green lines represent the
        overall attack rates of the population in three cities, predicted by the age-structured epidemic model \eqref{mainmodel}.}
    \label{fig:ar_three_cities}
\end{figure}

\newpage
 
\begin{table}[htbp]
\centering
\caption{Age-stratified epidemiological parameters in HFMD modeling}
\label{tab:hfmd_params}
\begin{tabular}{lcccc}
\toprule
Age group & Incubation period & Infectious period & Relative susceptibility & Relative infectivity \\
          & (days)            & (days)             & $u_i$                    & $v_j$                 \\
\midrule
{[}0,5)   & 4.4  & 7.0  & 1.00 & 1.00 \\
{[}5,10)  & 4.7  & 7.0  & 0.25 & 1.00 \\
{[}10,15) & 5.7  & 7.0  & 0.25 & 1.00 \\
{[}15,20) & 5.7  & 7.0  & 0.25 & 1.00 \\
{[}20,25) & 5.7  & 7.0  & 0.25 & 1.00 \\
{[}25,30) & 5.7  & 7.0  & 0.25 & 1.00 \\
{[}30,35) & 5.7  & 7.0  & 0.25 & 1.00 \\
{[}35,40) & 5.7  & 7.0  & 0.25 & 1.00 \\
{[}40,45) & 5.7  & 7.0  & 0.25 & 1.00 \\
{[}45,50) & 5.7  & 7.0  & 0.25 & 1.00 \\
{[}50,55) & 5.7  & 7.0  & 0.25 & 1.00 \\
{[}55,60) & 5.7  & 7.0  & 0.25 & 1.00 \\
{[}60,65) & 5.7  & 7.0  & 0.25 & 1.00 \\
{[}65,70) & 5.7  & 7.0  & 0.25 & 1.00 \\
70+       & 5.7  & 7.0  & 0.25 & 1.00 \\
\bottomrule
\multicolumn{5}{p{0.92\textwidth}}{\footnotesize \textit{Note:} 
Incubation periods are median values estimated from school
outbreaks in Hong Kong for kindergarten, primary and secondary school children ($4.4$, $4.7$ and $5.7$ days)~\cite{Yang2017hfmd}. Relative susceptibility $u_i$ is estimated from data in \cite{Li2013hfmd}. The infectious period ($7$ days) and relative
infectivity ($v_j=1$) are assumptions, taken as uniform across ages in the absence of age-specific evidence. The targeted effective reproduction number is $\mathcal{R}_0 = 1.5$, which is consistent with the value of range $1.4$-$1.6$ reported in \cite{Wang2011}. 
}
\end{tabular}
\end{table}

\newpage

\begin{figure}[htbp]
    \centering
    \includegraphics[width=0.8\linewidth]{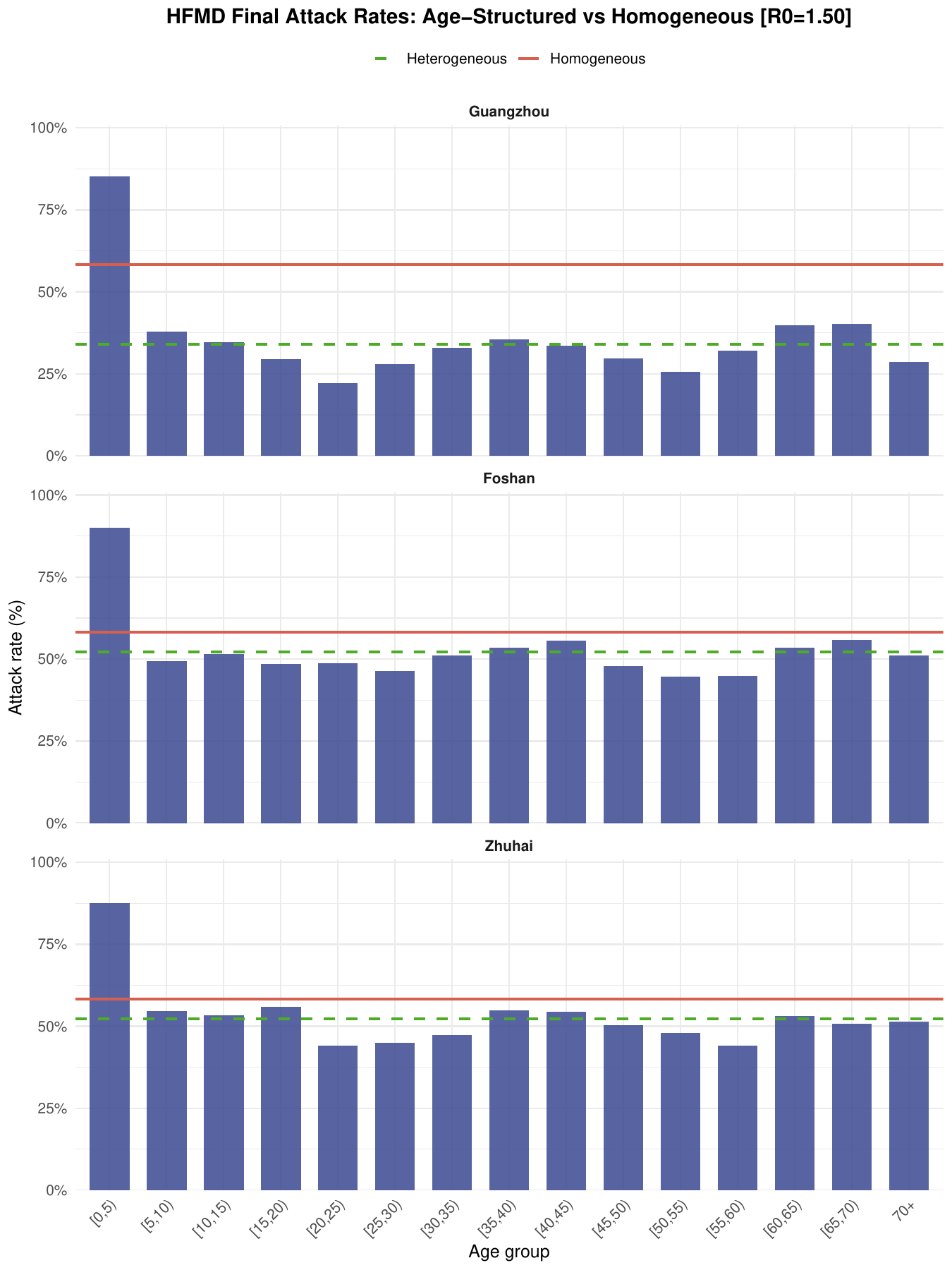}
    \caption{{Simulated HFMD final attack rates by age group across
        three cities in Guangdong Province (with assumed effective reproduction number $1.50$).}
        Bars show attack rates $\pi_i$ in all age groups obtained from the final-size relation in
        \eqref{final_size_relation}. The solid red horizontal line represents the overall attack
        rate predicted by the homogeneous mixing model with the identical $\mathcal{R}_0$ and population-weighted parameters. The dashed green lines represent the overall attack rates of the population in three cities, predicted by the age-structured epidemic model.} \label{fig:comparison_attack_rates_homogenous_heterogeneous_outbreak_outbreak_HFMD}
\end{figure}



\end{document}